\documentclass[
superscriptaddress,
nofootinbib,
twocolumn,
amsmath,amssymb,
aps,
pra,
notitlepage,
longbibliography
]{revtex4-1}

\usepackage{dcolumn}
\usepackage{bm}
\usepackage{epsfig}
\usepackage{graphicx}
\usepackage{latexsym}
\usepackage{amsfonts}
\usepackage{bigstrut}
\usepackage{setspace}
\usepackage{graphicx}
\usepackage{amsmath}
\usepackage{verbatim}
\usepackage{color}
\usepackage{SIunits}
\usepackage{hyperref}
\usepackage{multirow}
\usepackage{chemfig}
\usepackage{textgreek}

\newcommand{\be}{\begin{equation}}
\newcommand{\ee}{\end{equation}}
\newcommand{\ba}{\begin{array}}
\newcommand{\ea}{\end{array}}
\newcommand{\bqa}{\begin{eqnarray}}
\newcommand{\eqa}{\end{eqnarray}}

\renewcommand{\arraystretch}{1.5}

\begin{document}

\title{Optomechanical crystal with bound states in the continuum}

\author{Shengyan Liu} 
\thanks{These authors contributed equally to this work.}
\affiliation{Holonyak Micro and Nanotechnology Laboratory and Department of Electrical and Computer Engineering, University of Illinois at Urbana-Champaign, Urbana, IL 61801 USA}
\affiliation{Illinois Quantum Information Science and Technology Center, University of Illinois at Urbana-Champaign, Urbana, IL 61801 USA}
\author{Hao Tong} 
\thanks{These authors contributed equally to this work.}
\affiliation{Holonyak Micro and Nanotechnology Laboratory and Department of Electrical and Computer Engineering, University of Illinois at Urbana-Champaign, Urbana, IL 61801 USA}
\affiliation{Illinois Quantum Information Science and Technology Center, University of Illinois at Urbana-Champaign, Urbana, IL 61801 USA}
\author{Kejie Fang} 
\email{kfang3@illinois.edu}
\affiliation{Holonyak Micro and Nanotechnology Laboratory and Department of Electrical and Computer Engineering, University of Illinois at Urbana-Champaign, Urbana, IL 61801 USA}
\affiliation{Illinois Quantum Information Science and Technology Center, University of Illinois at Urbana-Champaign, Urbana, IL 61801 USA}

\begin{abstract} 
Chipscale micro- and nano-optomechanical systems, hinging on the intangible radiation-pressure force, have shown their unique strength in sensing, signal transduction, and exploration of quantum physics with mechanical resonators. Optomechanical crystals, as one of the leading device platforms, enable simultaneous molding of the band structure of optical photons and microwave phonons with strong optomechanical coupling. Here, we demonstrate a new breed of optomechanical crystals in two-dimensional slab-on-substrate structures empowered by mechanical bound states in the continuum (BICs) at 8 GHz. We show symmetry-induced BIC emergence with optomechanical couplings up to $g/2\pi\approx 2.5$ MHz per unit cell, on par with low-dimensional optomechanical crystals. Our work paves the way towards exploration of photon-phonon interaction beyond suspended microcavities, which might lead to new applications of optomechanics from phonon sensing to quantum transduction.
\end{abstract}

\maketitle
Cavity-optomechanics has attracted extensive studies in recent years because of the rich physics associated with the nonlinear optomechanical interaction and a broad range of prospective applications from signal transduction to sensing \cite{aspelmeyer2014cavity}. One of the leading optomechanical device architectures is optomechanical crystals \cite{eichenfield2009optomechanical}, where micro- and nano-scale structures give rise to strong radiation-pressure force coupling between wavelength-similar optical photons and microwave phonons. By band-structure engineering of suspended optomechanical crystals, both one-dimensional and quasi-two-dimensional defect cavities \cite{eichenfield2009optomechanical,safavi2014two,ren2020two} have been created with long-lived optical and mechanical resonances \cite{maccabe2020nano}. Such optomechanical crystal microcavities have enabled groundbreaking quantum experiments including ground-state cooling of mechanical resonators \cite{chan2011laser}, testing Bell inequality \cite{marinkovic2018optomechanical}, and a mechanical quantum memory \cite{wallucks2020quantum}.

Despite the success of optomechanical microcavities, it is highly desirable to explore two-dimensional optomechanical crystals. On one hand, two-dimensional optomechanical crystals offer more degrees of freedom for manipulation of photon-phonon interaction to induce collective phenomena \cite{ludwig2013quantum,brendel2017pseudomagnetic,brendel2018snowflake,ren2020topological}. On the other hand, extended optomechanical crystals, especially in unsuspended structures \cite{sarabalis2017release,qi2021nonsuspended,zhang2021silicon}, might alleviate the optical-absorption induced heating that plagues released microcavities \cite{meenehan2014silicon}. Ideally, such slab-on-substrate optomechanical crystals should facilitate dissipation of heat phonons while sustaining long-lived mechanical resonances in the device layer.

Recently, mechanical bound states in the continuum (BICs) are observed in two-dimensional slab-on-substrate phononic crystals \cite{tong2020observation}. Despite having a zero Bloch wavevector and thus spectrally immersing in the sound cone of the substrate, these mechanical BICs are confined in the slab because of the symmetry-induced decoupling from the acoustic radiation field. There are also proposals and demonstrations of mechanical BICs in microcavities due to, for example, accidental radiation amplitude cancellation \cite{chen2016mechanical,yu2021observation}.  A significant step forward thus would be coupling mechanical BICs with optical resonances in an optomechanical crystal which will bring the effective radiation-pressure force control and associated functionalities.

In this work, we realize two-dimensional silicon-on-insulator optomechanical crystals with mechanical BICs coupled with optical guided resonances. In such periodic optomechanical crystals, the radiation-pressure coupling between the mechanical BIC and optical modes strongly depends on the mode symmetry \cite{zhao2019mechanical}, which in many cases dictates an adversely null optomechanical coupling. Here, taking into account of both symmetry of the optomechanical crystal and the silicon crystal lattice, we are able to achieve optomechanical coupling up to $g/2\pi\approx 2.5$ MHz per unit cell between a mechanical BIC at 8 GHz and an optical band-edge mode at 193 THz, which is comparable to one-dimensional suspended optomechanical crystals \cite{chan2012optimized}. With optically-transduced mechanical spectroscopy at room temperature, we demonstrate control of mechanical BICs and optomechanical coupling via the interplay of symmetry of the optomechanical crystal and crystalline material. Our work paves the way for study of photon-phonon interaction in BIC optomechanical crystals at low temperatures, when the benefit of slab-on-substrate device architecture is expected to arise, and Floquet topological physics beyond the tight-binding model \cite{fang2019anomalous}.  \\

\noindent\textbf{Results}
\begin{figure*}
\centering
\includegraphics[width =\linewidth]{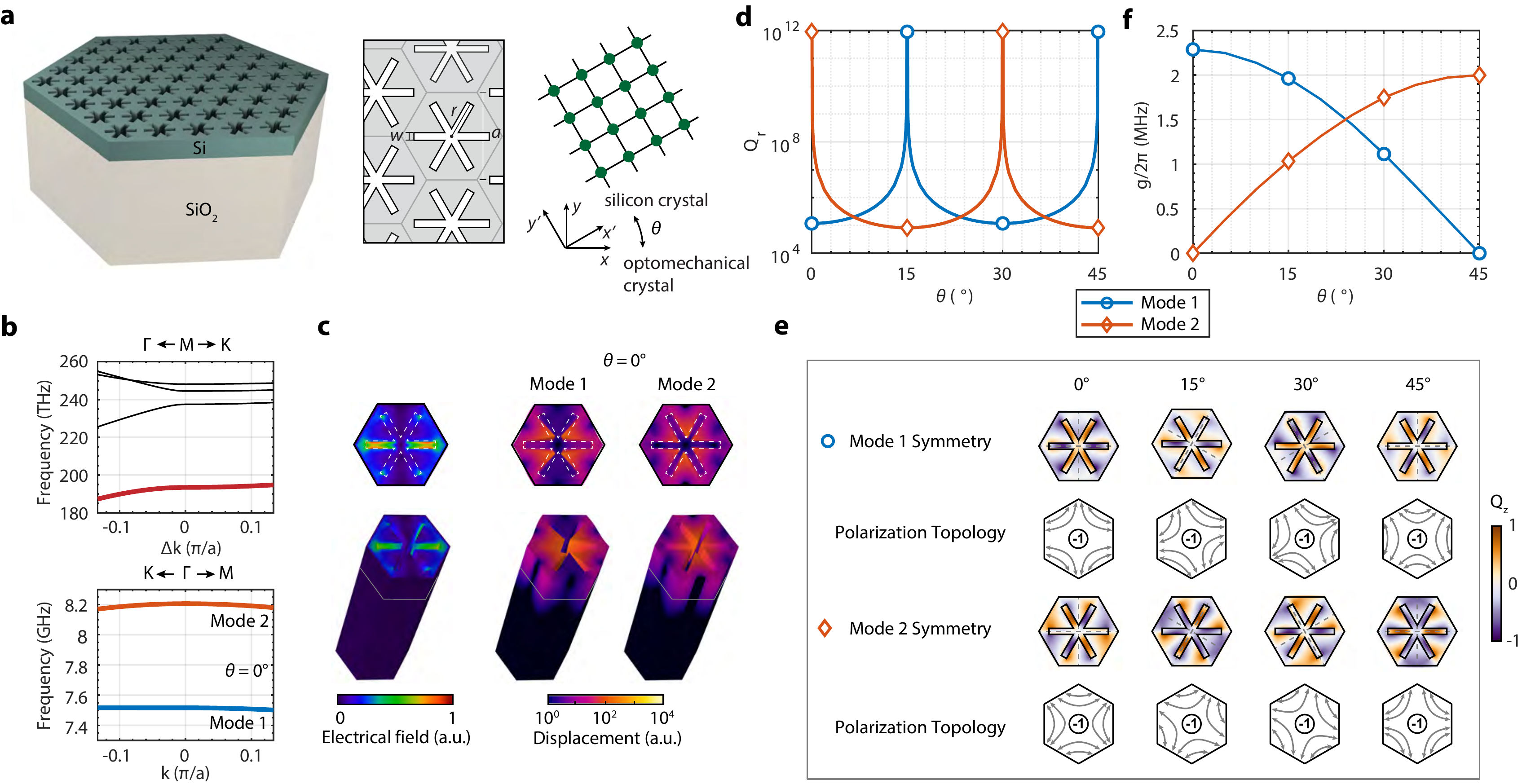}
\caption{\textbf{Two-dimensional optomechanical crystal with mechanical BICs.} \textbf{a}. A schematic diagram of the two-dimensional silicon-on-insulator optomechanical crystal and its unit-cell structure. The optomechanical crystal could be rotated by an angle $\theta$ relative to the silicon crystal lattice. \textbf{b}. Optical band structure near the $M$ point (top) and mechanical band structure near the $\Gamma$ point (bottom) for $\theta=0\degree$. The relevant optical and mechanical bands are highlighted in color. \textbf{c}. Simulated total electric field and total displacement field of the fundamental TE-like optical mode at the $M$ point and mechanical mode 1 and 2 at the $\Gamma$ point. \textbf{d}. Simulated radiative quality factor of the mechanical modes with respect to $\theta$. \textbf{e}. Mechanical mode symmetry illustrated using the $z$-direction displacement, momentum-space transverse polarization distribution and the winding number. \textbf{f}. Unit-cell optomechanical coupling of mechanical mode 1 and 2 with the fundamental TE-like optical mode.  }
\label{fig:1}
\end{figure*}

\noindent\textbf{Design of BIC optomechanical crystals.} The designed optomechanical crystal in silicon-on-oxide material system has a hexagonal ``snowflake'' unit cell (Fig. \ref{fig:1}a). The reason behind this design is that six-fold symmetric structures tend to yield mechanical BICs coupled with optical guided resonances \cite{zhao2019mechanical} and, also, the spike feature of the ``snowflake'' could lead to sizable vibrations for large optomechanical couplings. We consider optical modes at the $M$ point below the light cone and mechanical modes at the $\Gamma$ point. The latter is necessary because mechanical vibrations generally induce a linear perturbation of the optical mode energy, i.e., $\delta U\sim\int \delta x(\boldsymbol{R}) \epsilon|E(\boldsymbol{R})|^2d\boldsymbol{R}$, and this integral is nonzero in a periodic structure only when the mechanical displacement $\delta x$ has a zero Bloch wavevector (Supplementary Information (SI)).  Since the refractive index of crystalline silicon is isotropic, the symmetry of optical modes at the $M$ point is governed by the $C_{2v}$ group. We select the fundamental transverse-electric(TE)-like mode which is odd with respect to the $xz-$plane. The stiffness tensor of silicon, on the other hand, is anisotropic in the crystal plane with a $C_{4v}$ group symmetry, which is incommensurable with the $C_{6v}$ group symmetry of the hexagonal optomechanical crystal. As a result, the symmetry of mechanical modes at the $\Gamma$ point will depend on the orientation angle $\theta$ between the optomechanical crystal and the silicon crystal lattice. When $\theta = 0\degree$, $15\degree$, $30\degree$, and $45\degree$, the symmetry group of the mechanical mode is $C_{2v}$, while for other orientations it will be $C_2$. Only $C_{2v}$ group supports mechanical BICs which decouple from both transverse and longitudinal radiation waves \cite{tong2020observation}, whose displacement field is perpendicular and parallel to the wavevector, respectively.

Fig. \ref{fig:1}b shows the optical band structure near the $M$ point and mechanical band structure near the $\Gamma$ point for $\theta = 0\degree$ and $(r,w,a,t,h) = (167,34,389,220,3000)$ nm, where $t$ and $h$ are the thickness of the silicon and oxide layers, respectively, $a$ is the lattice constant, and $(r,w)$ are ``snowflake'' dimensions. The fundamental TE-like optical mode has a frequency of 193 THz and the relevant mechanical modes have frequencies about 7.5 and 8.2 GHz, respectively.  Their mode profiles are shown in Fig. \ref{fig:1}c. Simulation shows that the radiation quality factor $Q_r$, i.e., the ratio between the frequency and radiation loss rate, of mechanical mode 2 (8.2 GHz) diverges and that of mode 1 is finite but remains relatively high compared to other lossy modes. As a result, for $\theta = 0\degree$, mode 2 and 1 are mechanical BIC and quasi-BIC, respectively. The fact that mode 1 is a quasi-BIC is because the two mechanical bands are degenerate at the $\Gamma$ point when the stiffness tensor is isotropic and both modes are BICs belong to the $E_2$ representation of the $C_{6v}$ group; the actual anisotropic stiffness tensor of silicon splits the degeneracy, reducing the $E_2$ representation to $A_2$ (BIC) and $A_1$ (quasi-BIC) representations of the $C_{2v}$ group. Mode 1 turns out to have the $A_1$ representation, leading to coupling with the longitudinal radiation wave.  When $\theta$ changes, BIC and quasi-BIC alternate between mode 1 and 2 (while the frequencies of the two modes remain almost unchanged) as shown in Fig. \ref{fig:1}d. This is further illustrated in Fig. \ref{fig:1}e using the mode symmetry. Below, the mode symmetry under certain symmetry operation is defined with regard to the vector parity of the electric field of the optical mode or the displacement field of the mechanical mode. Taking $\theta = 0\degree$ as an example again, both mode 1 and 2 are even under the 180$\degree$ rotation, which leads to decoupling from the transverse radiation wave. In addition, mode 2(1) is odd(even) with respect to the symmetry axes (dashed lines), resulting decoupling(coupling) from/with the longitudinal radiation wave and thus a rigorous(quasi) BIC. Same arguments apply to the other three $\theta$'s while noticing the mirror symmetry axes rotate with $\theta$. The mechanical BIC and quasi-BIC are also associated with transverse topological charges, defined as the winding number of far-field transverse polarization around the $\Gamma$ point \cite{tong2020observation}. The polarization fields rotate together with $\theta$, which is unique for the anisotropic mechanical system. For orientations other than the four specific angles, the two modes belong to the $A$ representation of the $C_2$ group and thus are quasi-BICs which only couple to the longitudinal acoustic waves. As a result, their quality factor over $10^4$ is significantly higher than unconfined modes. 

The interaction between the optical and mechanical modes can be analyzed using the mode symmetry (SI). Roughly, because the $M$-point optical mode energy density is even with respect to the $xz-$plane, mechanical modes that are odd with respect to the $xz-$plane, including the BIC for $\theta=0\degree$ and $45\degree$, will not interact with the optical mode. For other cases, the optomechanical coupling could be nonzero (see Table \ref{tab:1} for a summary), thanks to the incommensurable symmetry of the optomechanical crystal and silicon lattice crystal. The bare optomechanical coupling of a unit cell, $g$, including both moving boundary and photoelastic effects, is calculated and plotted in Fig. \ref{fig:1}f. For example, the BIC at $\theta=15\degree$ has $g/2\pi=1.96$ MHz, with a contribution from the moving-boundary and photoelastic effect of  $0.61$ MHz and $1.35$ MHz, respectively. The coupling $g$ of the ``snowflake'' optomechanical crystal will increase with smaller air gaps. Here, with a practical air gap $w\approx 30$ nm, the BIC optomechanical crystal achieves a coupling rate (per unit cell) on par with one-dimensional suspended optomechanical crystals \cite{chan2012optimized}. 

\begin{table}[htbp]
	\centering
    \caption{\textbf{Summary of mechanical mode representation and optomechanical coupling.}}
    \scriptsize
	\begin{tabular}{|c|cccc|}
		\hline
		& $0\degree$     & $15\degree$    & $30\degree$    & $45\degree$ \\
		\hline
		Mode 1 & $A_1$, quasi-BIC & $A_2$, BIC   & $A_1$, quasi-BIC & $A_2$, BIC \\
		$g$    & $\ne 0$     & $\ne 0$     & $\ne 0$     & 0 \\
				\hline
		Mode 2 & $A_2$, BIC & $A_1$, quasi-BIC & $A_2$, BIC   & $A_1$, quasi-BIC \\
		$g$    & 0     & $\ne 0$     & $\ne 0$     & $\ne 0$ \\
		\hline
	\end{tabular}%
	\label{tab:1}
\end{table}

\begin{figure}[htb]
\centering
\includegraphics[width =\linewidth]{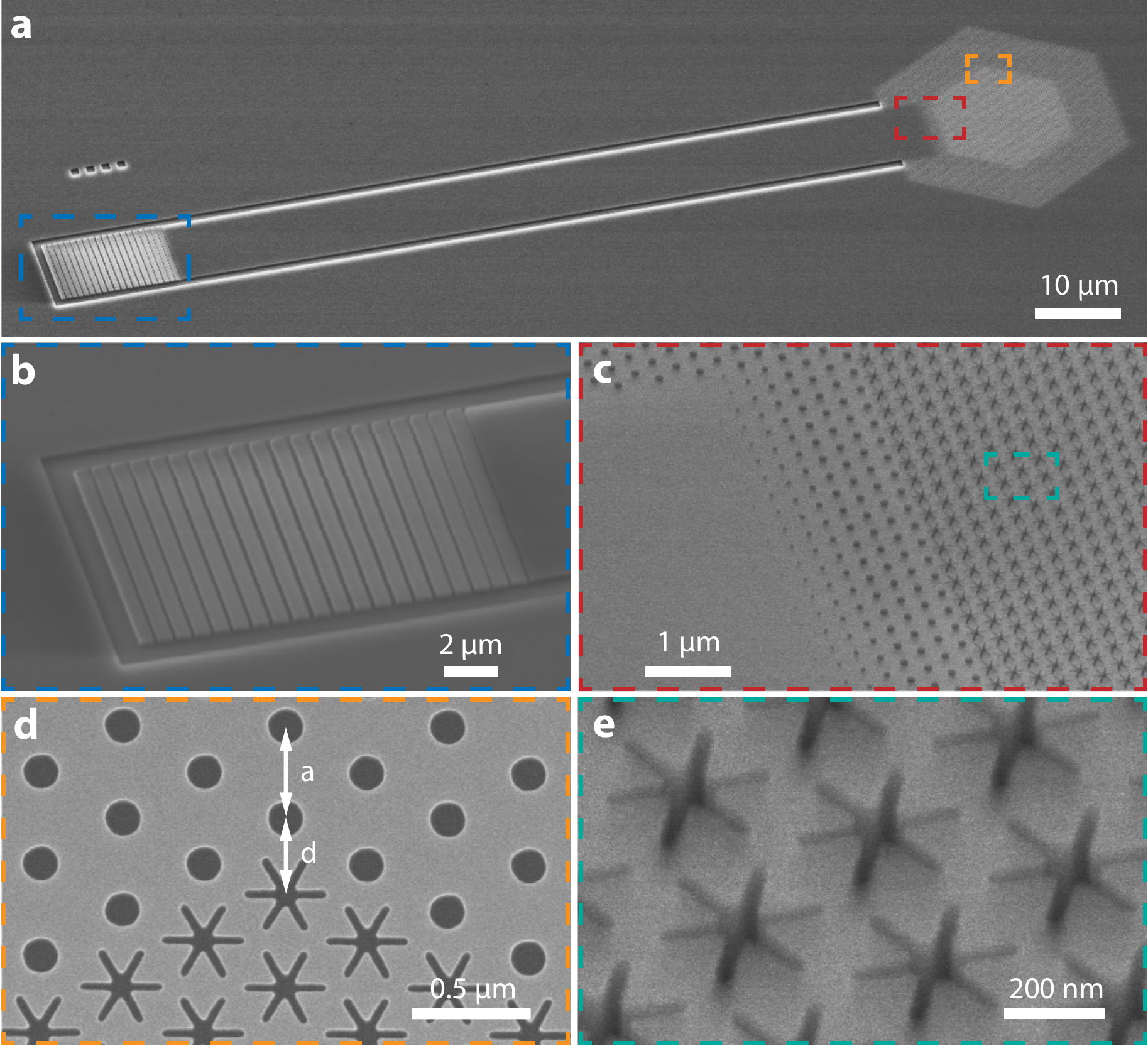}
\caption{\textbf{Scanning electron microscopy of the optomechanical crystal device.} \textbf{a}. Full device. \textbf{b}. Apodized grating coupler. \textbf{c}. Junction between the optomechanical crystal and waveguide with tapered photonic crystal mirrors. \textbf{d}. Boundary between the optomechanical crystal and photonic crystal mirror. The photonic crystal mirror is displaced by $d-a$. \textbf{e}. Zoom-in view of the optomechanical crystal. }
\label{fig:2}
\end{figure}

\begin{figure*}[htb]
\centering
\includegraphics[width =\linewidth]{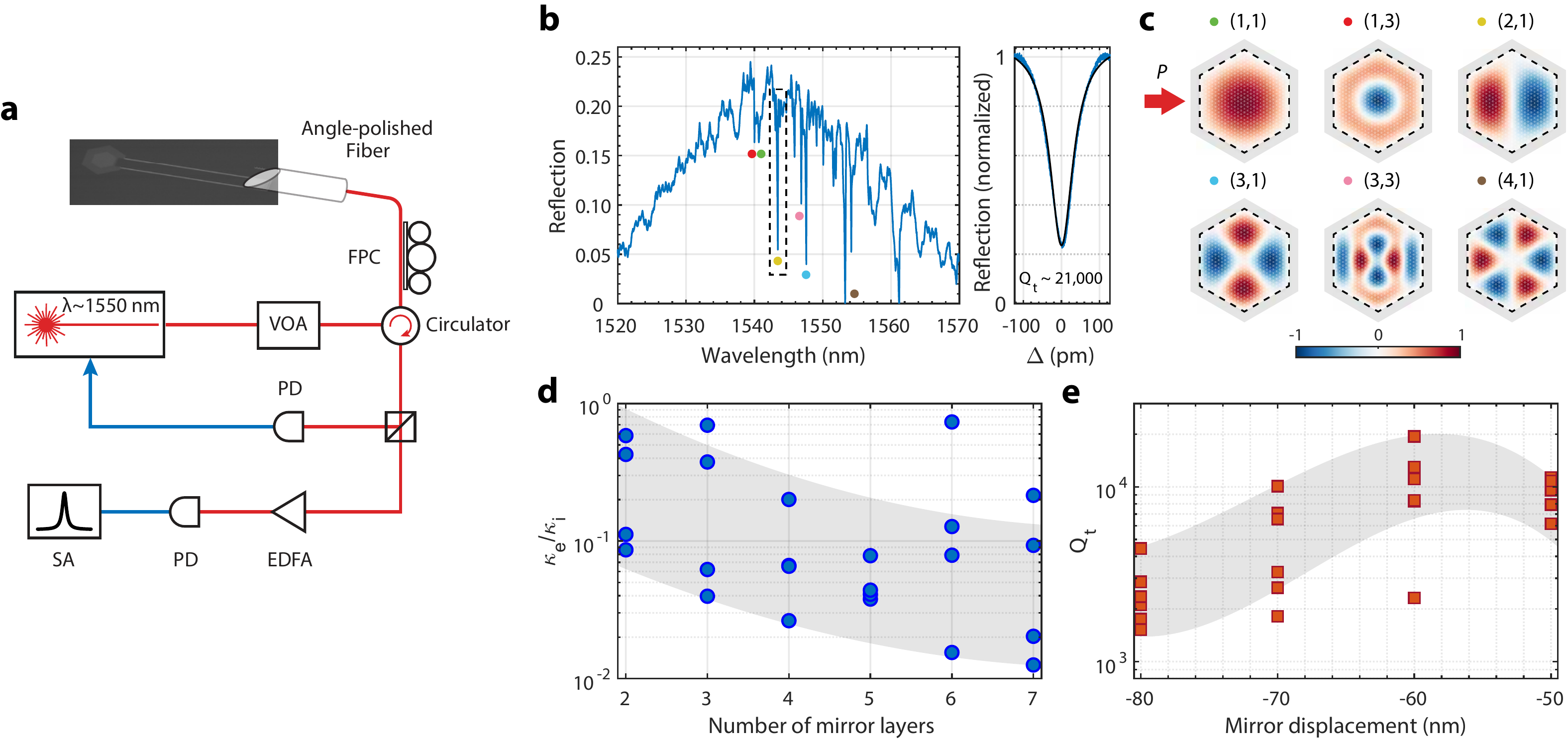}
\caption{\textbf{Multimode optical spectroscopy.} \textbf{a}. Schematic diagram of the measurement setup. The pump light is guided into the on-chip device using an angle-polished fiber. The reflected light is divided by a directional coupler for optical (10\%) and mechanical (90\%) spectroscopy. The optical signal is used to lock the laser-cavity detuning. VOA, variable optical attenuator. FPC, fiber polarization controller. EDFA, erbium-doped fiber amplifier. PD, photodetector. SA, spectrum analyzer. \textbf{b}. Optical spectrum of a device with multiple standing-wave resonances, one of which has a total quality factor of 21,000 (right). \textbf{c}. Mode envelop of standing-wave resonances with given orders corresponding to the dotted resonances in \textbf{b}. The arrow indicates the waveguide input. \textbf{d}. $\kappa_e/\kappa_i$ for different number of nominal mirror layers at the junction between the optomechanical crystal and waveguide. For each number of mirror layers, multiple devices with different photonic crystal mirror displacement are measured. \textbf{e}. Total optical quality factor versus the displacement of the photonic crystal mirror, i.e., $d-a$. \textbf{e} is plotted using the same set of data of \textbf{d}. }
\label{fig:3}
\end{figure*}

\noindent\textbf{Photonic crystal band-edge multimodes.} The optomechanical crystal device is fabricated from a silicon-on-insulator wafer with a 220 nm thick silicon device layer and a 3 \textmu m buried oxide. The air gap of the ``snowflake'' as small as 30 nm is achieved. The device consists of several functional components (Fig. \ref{fig:2}). The hexagonal BIC optomechanical crystal has $N$ unit cells along each edge and is surrounded by photonic crystal mirrors with a triangular lattice of cylindrical holes on five edges. The photonic crystal mirror, with the same lattice constant as the snowflake optomechanical crystal, is designed to have a complete TE-like bandgap with the center wavelength around 1550 nm to suppress the lateral radiation of the optical band-edge mode. The photonic crystal mirror is also slightly displaced to minimize the out-of-plane radiation due to the boundary effect on the finite-size optical mode (Fig. \ref{fig:2}d). One side of the hexagonal snowflake crystal is connected to a waveguide terminated with an apodized grating coupler for coupling light from a single-mode optical fiber. This configuration makes the optomechanical crystal effectively a one-port device with (mode-dependent) external, i.e., to-waveguide, and intrinsic loss rate of $\kappa_e$ and $\kappa_i$, respectively. The former can be controlled by the number of photonic crystal mirror layers between the waveguide and optomechanical crystal (Fig. \ref{fig:2}c) to achieve different coupling conditions including the critical coupling, i.e., $\kappa_e\approx\kappa_i$, which is desirable for sideband-resolved mechanical spectroscopy. 

The device is measured using the setup shown in Fig. \ref{fig:3}a. An angle-polished optical fiber is used to guide light via the on-chip apodized grating coupler \cite{snyder2013packaging,li2014silicon} to the optomechanical crystal and collect the reflected light. The angle of the fiber ($\approx 35.5\degree$) and the apodized grating coupler are co-designed to realize an optimized coupling efficiency of $52\%$ for 1550 nm light and 3-dB bandwidth of $40$ nm (SI). Because of the reflection at the boundary of finite-size photonic/phononic crystals, band-edge standing-wave resonances will be formed \cite{hood2016atom, jin2019topologically,tong2020observation,chua2014larger,chen2022analytical}. The mode envelope of the standing-wave resonances can be approximated by the $(p,q)-$th order eigenfunctions of a flat-top potential well within the boundary of the photonic/phononic crystal \cite{xu2005confined}. Fig. \ref{fig:3}b shows the optical reflection spectrum of a hexagonal optomechanical crystal with $N=25$, where a series of band-edge standing-wave resonances are observed. The major order $p$ of a standing-wave resonance is identified from the group of resonances it belongs to and the minor order $q$ is determined by the position of the resonance in a group. Because of the hyperbolic paraboloid topology of the band structure near the $M$ point, resonances with smaller $p$ and larger $q$ will have shorter wavelengths. In addition, only $q$-odd resonances are observed because the excited waveguide mode is even with respect to the center of the waveguide. Order $q=1$ modes are expected to have deepest resonance dip given they have largest $\kappa_e$ close to the critical coupling.  The mode envelop of some standing-wave resonances are shown in Fig. \ref{fig:3}c.

\begin{figure*}[htb]
\centering
\includegraphics[width =1\linewidth]{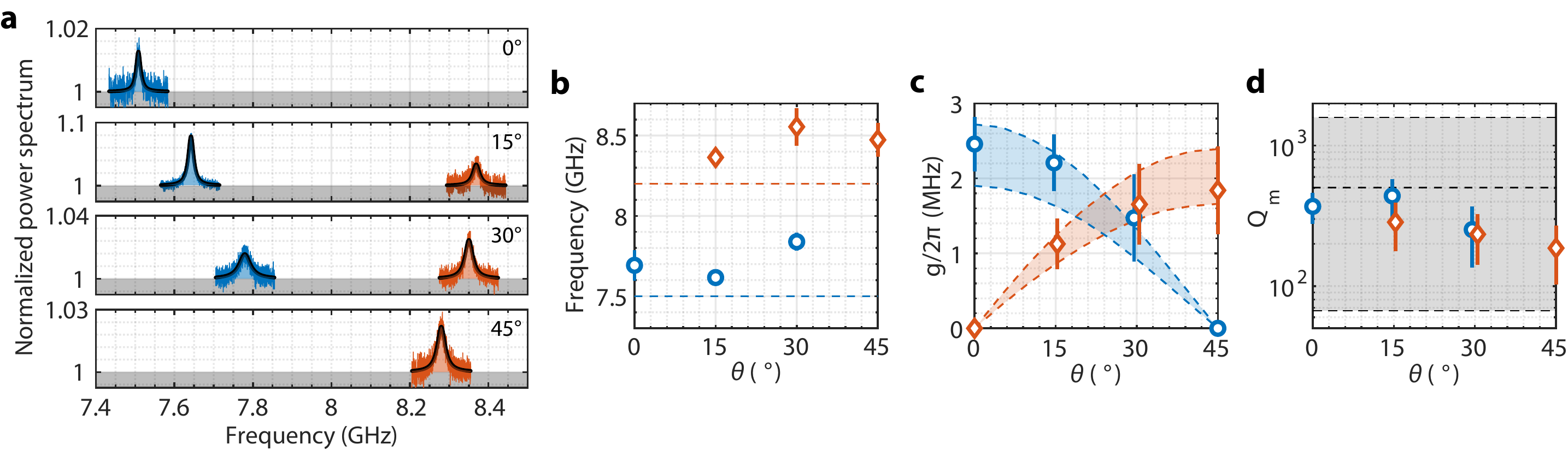}
\caption{\textbf{Room-temperature mechanical spectroscopy.} \textbf{a}. Measured mechanical noise power spectrum for one device each for $\theta=0\degree$, $15\degree$, $30\degree$ and $45\degree$. \textbf{b}. Mechanical frequency distribution of a group of five devices for each $\theta$. Error bar is the standard deviation. Dashed lines are the simulated mechanical frequencies.  \textbf{c}. Unit-cell optomechanical coupling. The dashed lines correspond to simulated coupling for $w = 28$ nm (top) and $40$ nm (bottom). \textbf{d}. Total quality factor of the mechanical modes. The dashed lines are the estimated bound and mean from the simulation of  a $1\times 4$ super-cell with disorders. }
\label{fig:4}
\end{figure*}

The number of photonic crystal mirror layers at the waveguide-optomechanical crystal junction is varied to tune $\kappa_e/\kappa_i$, which is extracted by fitting the optical resonance spectrum using the normalized reflection coefficient of the one-port waveguide-coupled resonator, $R[\omega]=\left|1-\frac{\kappa_e}{i(\omega-\omega_o)+(\kappa_e+\kappa_i)/2} \right|^2$. Fig. \ref{fig:3}d shows $\kappa_e/\kappa_i$ of the $(1,1)$ standing-wave resonance of a group of devices, where a decreasing trend over the number of junction mirror layers is observed as a result of the reduced coupling between the waveguide mode and standing-wave resonances. The displacement of the photonic crystal mirror around the optomechanical crystal is also varied to optimize the total quality factor of the standing-wave resonance. As shown in Fig. \ref{fig:3}e, a displacement of $-60$ nm from the nominal lattice constant is optimal, which is consistent with the simulation (SI). We note the variation of photonic crystal mirrors not only changes $\kappa_e$ but also $\kappa_i$ via the perturbation of the evanescent field, leading to fluctuations of $\kappa_e/\kappa_i$. An optimized $(2,1)$ optical resonance is shown in Fig. \ref{fig:3}b with a total quality factor of $21,000$ while being close to the critical coupling. The quality factor of the optical resonance and its variation is largely due to the disorder induced scattering to the modes above the light cone \cite{jin2019topologically}.

\noindent\textbf{Room-temperature mechanical spectroscopy.}
We performed the mechanical spectroscopy at room temperature using a blue-detuned laser with a frequency $\omega_l=\omega_o+\omega_m$, where $\omega_o$ and $\omega_m$ are the frequencies of the optical and mechanical band-edge modes, respectively. We stabilized the laser-cavity-detuning by locking the device-reflected power with feedback control of the laser frequency (Fig. \ref{fig:3}a). For each device, an optical standing-wave resonance with $Q_t$ about $2\times 10^4$ and $\kappa_e/\kappa_i$ close to 1 is chosen,
which means the device is operated near the sideband-resolved regime, i.e., $\omega_m\approx \omega_o/Q_t$. The reflected pump light with sidebands due to the modulation of mechanical modes is sent to a high-speed photodetector. The beating between the pump and sidebands thus yields the mechanical spectrum which is observed by a spectrum analyzer. 

Devices with different orientations between the optomechanical crystal and the silicon crystal lattice are fabricated to reveal the mechanical (quasi-)BICs and the impact of symmetry on the optomechanical interaction. Fig. \ref{fig:4}a shows the mechanical spectroscopy of a group of devices with $\theta=0\degree$, $15\degree$, $30\degree$, and $45\degree$. The spectrum is normalized with respect to the total background noise.  At $\theta=0\degree$ and $45\degree$, only the quasi-BIC mode is observed while the BIC mode is invisible because of the symmetry-inhibited optomechanical coupling.  At $\theta=15\degree$ and $30\degree$, both BIC and quasi-BIC modes are detected. These observed mechanical modes are the (1,1) standing-wave resonance, while higher order resonances are obscure because of both lower quality factor and optomechanical coupling. Fig. \ref{fig:4}b shows the frequency distribution of a group of devices. The occurrence of the mechanical BIC and quasi-BIC modes and their frequencies are consistent with the simulation. 

Under a blue-detuned pump, the optomechanical interaction between the mechanical and optical resonances is described by a linearized two-mode squeezing Hamiltonian, $\hat H=G(\hat a^\dagger \hat b^\dagger+\hat a\hat b)$, where $\hat a^\dagger$($\hat b^\dagger$) and $\hat a$($\hat b$) are the creation and annihilation operators of the optical(mechanical)  resonance. The parametrically-enhanced optomechanical mode coupling $G$ of standing-wave resonances is given by 
\be\label{G}
G=\zeta g \sqrt{n_c},
\ee
where $n_c$ is the average number of photons in one unit cell and $\zeta$ is a resonance-dependent $O(1)$ parameter due to finite-mode-size correction (SI). 
Given the nature of weak radiation-pressure force, the important coupling in generic optomechanical systems is the parametric mode coupling, which is related to the cooperativity given by $C=4G^2/\kappa\gamma$.
It is seen from Eq. \ref{G} that the parametric mode coupling of a finite optomechanical crystal is primarily determined by the unit-cell coupling and the per-unit-cell photon number, and is independent of the size of the crystal. In other words, despite that the bare mode coupling of optomechanical crystals roughly scales as $1/N$, it can be compensated by the large number of photons available which scales with $N^2$.  The per-unit-cell photon number is largely constrained by the thermo-optic effect and the heat capacity of the unit cell, which the slab-on-substrate structure can ameliorate, especially when the thermal conductivity of the substrate is comparable to the slab. In our experiments, the device is typically operated under a pump power $P$ of a few mW and $n_c=\kappa_eP/((\omega_m)^2+(\kappa/2)^2)/\frac{3\sqrt{3}}{2}N^2$ is on the order of $10-20$.
The unit-cell optomechanical coupling $g$ is extracted from the noise power spectrum (see Methods and SI) and plotted in Fig. \ref{fig:4}c. Large optomechanical coupling $g/2\pi\approx 2.5$ MHz is observed for the mechanical BIC. The deviation from the simulated value could be due to the variation of the actual snowflake gap size and disorders in the optomechanical crystal. Despite being unreleased and two-dimensional, the optomechanical interaction (per unit cell) in the BIC optomechanical crystal is on par with the suspended low-dimensional optomechanical crystal devices, such as nanobeams. 

The quality factor of the observed mechanical modes is shown in Fig. \ref{fig:4}d. There are two main factors limiting the mechanical quality factor of the current device. First, the size of the optomechanical crystal is $N=25$, corresponding to a finite wavevector of $k\approx 2\pi/(3Na)$ for the fundamental standing-wave resonance. The radiation quality factor of mechanical standing-wave resonances with finite wavevectors decreases drastically as they deviate from the BIC at the $\Gamma$ point \cite{tong2020observation}. Second, the fabricated optomechanical crystal has critical dimensions as small as 30 nm, which cause random variations among unit cells. These disorders induce scattering between different orders of band-edge resonances, which effectively introduces more radiation channels to a given resonance and degrades its quality factor \cite{regan2016direct, ni2017analytical, jin2019topologically}. The lateral radiation for this size of crystal is not a dominant loss according to the simulation and can be optimized with a phononic crystal mirror.  In addition, at room temperature, the quality factor of GHz-frequency mechanical modes is ultimately limited by material absorption to around 1000-2000 \cite{eichenfield2009optomechanical}. Because of these, there is no significant statistical difference of the quality factor of BIC and quasi-BIC modes. The observed quality factor is verified with numerical simulations of finite super-cells with realistic disorders (SI). However, the mechanical radiation quality factor will increase with the size of optomechanical crystals \cite{tong2020observation}, which will be critical to low-temperature measurements when the material absorption is suppressed.

\noindent\textbf{Discussion}\\
In summary, we have realized the first two-dimensional slab-on-substrate optomechanical crystals with mechanical BICs. Guided by a symmetry-based design approach, this architecture offers optomechanical interaction (per unit cell) on par with suspended optomechanical crystal devices. The two-dimensional optomechanical crystal with tunable symmetry provides an arena for exploration of rich multimode physics \cite{nielsen2017multimode,renninger2018bulk}. In addition, the cavity-less optomechanical crystal might realize Floquet topological physics beyond the tight-binding model \cite{fang2019anomalous}. The benefit of the slab-on-substrate device architecture in terms of heat dissipation is obscured at room temperature, especially for optomechanical crystals with a large area-to-perimeter ratio, because of the low thermal conductivity of the oxide substrate two orders smaller than silicon. However, such benefit will become evident at low temperatures ($<1$ K), relevant for quantum experiments involving GHz mechanical modes, or in a different material systems, when/where the thermal conductivity of the substrate and slab becomes comparable \cite{zeller1971thermal,thompson1961thermal}. Besides the suppressed material-absorption loss at low temperatures, the mechanical quality factor could be enhanced using the merging BIC mechanism \cite{jin2019topologically} and implementing appropriate phononic crystal mirrors. As a consequence, slab-on-substrate BIC optomechanical crystals with improved optical and mechanical losses could be unique at low temperatures for exploration of modalities including phonon sensing and macroscopic mechanical oscillators in the quantum regime \cite{chu2017quantum,kotler2021direct}.

\noindent\textbf{Methods}\\
\textbf{Device fabrication.} Devices are fabricated in silicon-on-insulator microchips (220 nm silicon device layer and 3 $\upmu$m buried oxide layer) using electron beam lithography with ZEP520A mask, followed by inductively coupled plasma reactive ion etch of silicon using \chemfig{SF_6} and \chemfig{CHF_3}. \\ 
\textbf{Mechanical noise power spectrum.} The total noise power spectral density measured by the photodetector is given by \cite{safavi2013laser,meenehan2014silicon}
\be\label{mainspectrum}
    \begin{aligned}
        S[\omega] &= {S_e}+\frac{G_e^2}{R_I}\left(  {S_{\mathrm{EDFA}}} + G^2_{\mathrm{EDFA}}{S^2_{\mathrm{SN}}}{\Big(1 + \eta \frac{{{\kappa _e}}}{\kappa }\frac{{8{G^2}}}{\kappa }{{S}_m}[\omega]\Big)} \right)
    \end{aligned}
\ee
where $S_e$ is the electronic noise of the detector, $S_{\mathrm{EDFA}}$ is the noise of EDFA, $S_{\mathrm{SN}}=\sqrt{2P_{\mathrm{out}}\hbar\omega_l}$ is the optical shot noise, $S_m[\omega]$ is the mechanical noise spectrum, $\kappa=\kappa_i+\kappa_e$ is the total loss rate of the optical resonance, $G$ is the parametrically-enhanced optomechanical coupling, $\eta$ is the total detection efficiency, $G_{\mathrm{EDFA}}$ is the EDFA gain, $G_e$ is the detector gain factor from optical power to voltage, and $R_I$ is the input impedance of the spectrum analyzer. The optically-transduced mechanical noise spectrum is given by
\begin{equation}\label{mechspectrum}
    {S_m}[\omega] = \frac{1}{2}\left(\frac{{\gamma (\bar n_m + 1)}}{{{{\left( {{\omega _m} - \omega } \right)}^2} + {{\left( {\gamma /2} \right)}^2}}}+\frac{{\gamma \bar n_m}}{{{{\left( {{\omega _m} +\omega } \right)}^2} + {{\left( {\gamma /2} \right)}^2}}}\right)
\end{equation}
with $\bar n_m=\frac{k_BT}{\hbar\omega_m}$ the thermal occupation of the mechanical mode. The total detection efficiency is $\eta=\eta_{\mathrm{cpl}}\eta_{\mathrm{t}}\eta_{\mathrm{det}}$, where $\eta_\mathrm{cpl}$ is the coupling efficiency of the grating coupler at the pump wavelength, $\eta_{\mathrm{t}}$ is the total transmission efficiency in the optical fiber path from the chip to the detector, and $\eta_{\mathrm{det}}$ is the quantum efficiency of the photodetector. We measured $\eta_\mathrm{t}=0.80$ and $\eta_\mathrm{cpl}\approx 0.5$ depending on the pump wavelength, while $\eta_\mathrm{det}=0.68$ is given by the photodetector, which results in  $\eta\approx 0.27$. ${S_e}$ is determined by blocking the light and $\frac{G_e^2}{R_I}G^2_{\mathrm{EDFA}}{S^2_{\mathrm{SN}}}$ is determined by removing EDFA while keeping the optical power incident onto the photodetector the same. Then the measured noise power spectrum is fitted using Eqs. \ref{mainspectrum} and \ref{mechspectrum}, with $\omega_m$, $\gamma$, and $G$ the only fitting parameters. The unit-cell optomechanical coupling is calculated from $g=G/(\zeta \sqrt{n_c})$, using $n_c=\kappa_eP/((\omega_m)^2+(\kappa/2)^2)/\frac{3\sqrt{3}}{2}N^2$ and $\zeta$ depending on the optical standing-wave resonance that is used for the mechanical spectroscopy (SI).

\vspace{2mm}
\noindent\textbf{Acknowledgements}\\ 
This work is supported by US National Science Foundation (Grant No. 1944728 and 2016136) and Office of Naval Research (Grant No. N00014-21-1-2136).

\onecolumngrid
\appendix

\section{Symmetry analysis of mechanical BICs and optomechanical coupling}

\begin{figure}[htb]
	\centering
	\includegraphics[width=0.5\linewidth]{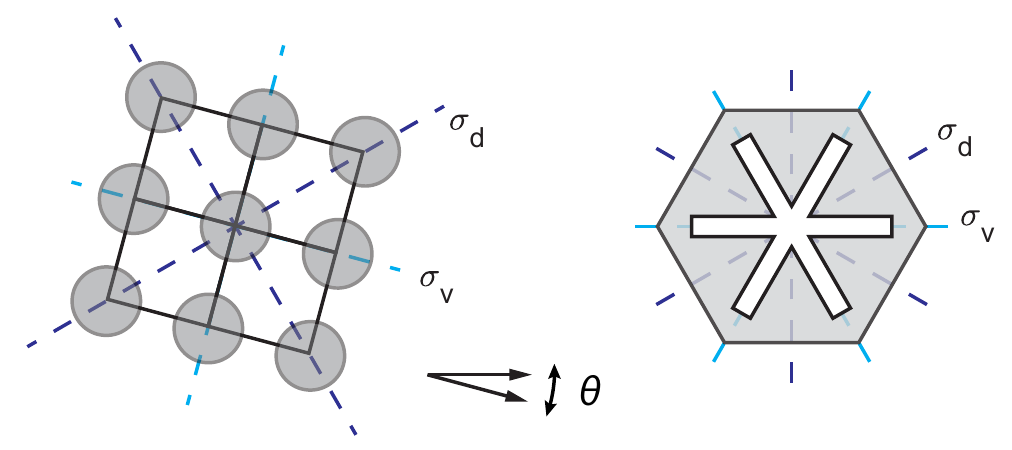}
	\caption{Schematic illustration of the symmetry of the silicon crystal plane and optomechanical crystal.}
	\label{fig:figs1v1}
\end{figure}

The symmetry group of the stiffness tensor of silicon in the crystal plane and the hexagonal optomechanical crystal is $C_{4v}$ and $C_{6v}$, respectively. For both $C_{4v}$ and $C_{6v}$ groups, there are two sets of equivalent mirror planes $\sigma_v$ and $\sigma_d$ as shown in Fig. \ref{fig:figs1v1}. When one mirror plane of the silicon crystal lattice aligns with one mirror plane of the hexagonal optomechanical crystal, i.e., $\theta = 0\degree$, $15\degree$, $30\degree$, and $45\degree$, the mechanical system restores mirror symmetry and is described by $C_{2v}$ group. Otherwise, the symmetry group for the mechanical system is $C_2$. 

The far-field acoustic plane wave propagating along the $z-$direction can be written as
\begin{equation}
	\bm{Q}_{0}=\bm{Q}_{T, 0}+\bm{Q}_{L, 0}=\left(u \bm{e}_{x}+v \bm{e}_{y}\right) e^{i k_{T, z}^{0} z}+w \bm{e}_{z} e^{i k_{L, z}^{0}z}.
\end{equation}
According to Tables \ref{tab:S1} and \ref{tab:S2}, $\bm{e}_{x}$ and $\bm{e}_{y}$ belong to $B_1$ and $B_2$ representations of $C_{2v}$ and $B$ representation of $C_2$, respectively, and $\bm{e}_{z}$ belongs to $A_1$ representation of $C_{2v}$ and $A$ representation of $C_2$. Therefore, a mechanical mode at the $\Gamma$ point can be a BIC, i.e., decouples from both transverse and longitudinal radiation waves, only if it belongs to the $A_2$ representation of $C_{2v}$ group when $\theta = 0\degree$, $15\degree$, $30\degree$, or $45\degree$. 
\begin{table}[htbp]
	\centering
	\caption{Character table of the $C_{2v}$ point group}
	\begin{tabular}{|c|cccc|}
		\hline
		$C_{2v}$   & $E$     & $C_2$    & $\sigma_v$ & $\sigma_d$ \\
		\hline
		$A_1$    & 1     & 1     & 1     & 1     \\
		$A_2$    & 1     & 1     & -1    & -1     \\
		$B_1$    & 1     & -1    & 1     & -1    \\
		$B_2$    & 1     & -1    & -1    & 1    \\
		\hline
	\end{tabular}%
	\label{tab:S1}%
\end{table}%

\begin{table}[htbp]
	\centering
	\caption{Character table of the $C_{2}$ point group}
	\begin{tabular}{|c|cc|}
		\hline
		$C_2$    & $E$     & $C_2$   \\
		\hline
		$A$     & 1     & 1    \\
		$B$     & 1     & -1   \\
		\hline
	\end{tabular}%
	\label{tab:S2}%
\end{table}%

Next we analyze the optomechanical coupling of a unit cell based on the symmetry. The unit-cell optomechanical coupling is given by
\begin{equation}
	g= {g_{{\mathrm{MB}}}} + {g_{{\mathrm{PE}}}} \equiv \sqrt {\frac{\hbar }{{2{m_{{\mathrm{eff}}}}{\omega _m}}}} \left( {{g_{{\mathrm{OM,MB}}}} + {g_{{\mathrm{OM,PE}}}}} \right)
\end{equation}
where $m_{\mathrm{eff}}=\int_{\mathrm{u.c.}} \rho|\bm{Q}|^{2} \mathrm{~d} V$ is the unit-cell effective mass, $\omega_m$ is the mechanical frequency. The moving-boundary and photoelastic components are calculated as
\begin{equation}\label{gmb}
	g_{\mathrm{OM}, \mathrm{MB}}=-\frac{\omega_{o}}{2} \frac{\int_{\mathrm{u.c.}}(\bm{Q} \cdot \hat{\bm{n}})\left(\Delta \epsilon\left|\bm{E}_{\|}\right|^{2}-\Delta \epsilon^{-1}\left|\bm{D}_{\perp}\right|^{2}\right) \mathrm{~d} S}{\int_{\mathrm{u.c.}} \bm{E}^{*} \cdot \bm{D} \mathrm{~d} V},
\end{equation}
and
\begin{equation}\label{gpe}
	g_{\mathrm{OM}, \mathrm{PE}}=\frac{\omega_{o}}{2} \frac{\int_{\mathrm{u.c.}} \epsilon_{0} n^{4} E_{i}^{*} E_{j} p_{i j k l} S_{k l} \mathrm{~d} V}{\int_{\mathrm{u.c.}} \bm{E}^{*} \cdot \bm{D} \mathrm{~d} V},
\end{equation}
where $\omega_{o}$ is the optical frequency, $\hat{\bm{n}}$ is the normal vector of the interfaces, $\bm{Q}$ is displacement, $S_{kl}$ is strain tensor, the subscripts $\|$ and $\perp$ indicate the field components parallel and perpendicular to the surface and $\Delta \epsilon=\epsilon_{\mathrm{int }}-\epsilon_{\mathrm{ext}}$ and $\Delta \epsilon^{-1}=\epsilon_{\mathrm{int}}^{-1}-\epsilon_{\mathrm{ext}}^{-1}$ ($\epsilon_{\mathrm{ext}}$ is the permittivity of the media which $\hat{\bm{n}}$ points to and $\epsilon_{\mathrm{int}}$ is the permittivity of the media on the other side). 

For the moving boundary term, we write the numerator using a shorthanded notation $\int_{\mathrm{u.c.}}(\bm{Q} \cdot \hat{\bm{n}}) f \mathrm{~d} S$. When there is mirror symmetry, this integration can be calculated as
\begin{equation}
	\begin{aligned}
		\int_{\mathrm{u.c.}} {\bm{Q}\cdot\hat{\bm{n}}f\mathrm{~d}S}  &= \left( {\int_{\mathrm{u.c.}} {\bm{Q}(\bm{r})\cdot\hat{\bm{n}}(\bm{r})f(\bm{r})\mathrm{~d}S}  + \int_{\mathrm{u.c.}} {\bm{Q}(\sigma \bm{r})\cdot\hat{\bm{n}}(\sigma \bm{r})f(\sigma \bm{r})\mathrm{~d}S} } \right)/2\\
		&= \left( {\int_{\mathrm{u.c.}} {\bm{Q}(\bm{r})\cdot\hat{\bm{n}}(\bm{r})f(\bm{r})\mathrm{~d}S}  + \int_{\mathrm{u.c.}} {{\sigma ^{ - 1}}\bm{Q}(\bm{r})\cdot\hat{\bm{n}}(\bm{r}){\sigma ^{ - 1}}f(\bm{r})\mathrm{~d}S} } \right)/2\\
		&= (1 + {\chi _{\bm{Q}}}(\sigma ))\int_{\mathrm{u.c.}} 	{\bm{Q}(\bm{r})\cdot\hat{\bm{n}}(\bm{r})f(\bm{r})\mathrm{~d}S} /2,
	\end{aligned}
\end{equation}
where $\chi_{\bm{Q}}(\sigma)$ stands for the character of mirror symmetry operation $\sigma$. Thus, the integral vanishes when the mechanical mode is odd with respect to the mirror plane. We have already used the fact that $f(\bm{r})$ is even under the mirror operation as it is a quadratic function of $\bm{E}$ and $\bm{D}$. This derivation is valid only when both mechanical and optical modes share the same mirror plane, which is true for $\theta = 0\degree$ and $\theta = 45\degree$. For $\theta = 15\degree$ and $\theta = 30\degree$, $g_{\mathrm{OM}, \mathrm{MB}}$ in general will be nonzero.

To calculate the photoelastic term, we note the photoelastic tensor for cubic crystal, such as silicon, in Voigt notation is given by
\begin{equation}
	p_{\mathrm{C}}=\left(\begin{array}{cccccc}p_{11} & p_{12} & p_{12} & & & \\ & p_{11} & p_{12} & & & \\ & & p_{11} & & & \\ & & & p_{44} & & \\ & \mathrm{symm.} & & & p_{44} & \\ & & & & & p_{44}\end{array}\right).
\end{equation}
We perform the computation in the frame of optomechanical crystal by rotating the silicon crystal lattice. Under a rotating $R(\theta)$, the photoelastic tensor transforms as
$p_{i j k l}^{\prime}(\theta)=R_{i p}(\theta) R_{j q}(\theta) R_{k r}(\theta) R_{l s}(\theta) p_{p q r s}$, or in the Voigt notation,
\begin{equation}
\begin{array}{*{20}{c}}
	{{{p}_{11}^\prime} = {{p}_{22}^\prime} = \frac{1}{4}\left[ {{p_{11}}(\cos 4\theta  + 3) + \left( {{p_{12}} + 2{p_{44}}} \right)(1 - \cos 4\theta )} \right]}\\
	\begin{array}{c}
		{{p}_{33}^\prime} = {p_{33}}\\
		{{p}_{12}^\prime} = {{p}_{21}^\prime} = \frac{1}{4}\left[ {{p_{12}}(\cos 4\theta  + 3) + \left( {{p_{11}} - 2{p_{44}}} \right)(1 - \cos 4\theta )} \right]
	\end{array}\\
	\begin{array}{c}
		{{p}_{13}^\prime} = {{p}_{23}^\prime} = {{p}_{31}^\prime} = {{p}_{32}^\prime} = {p_{12}}\\
	\end{array}\\
	{{{p}_{44}^\prime} = {{p}_{55}^\prime} = {p_{44}}}\\
	{{{p}_{66}^\prime} = \frac{1}{4}\left[ {2{p_{44}}(\cos 4\theta  + 1) + \left( {{p_{11}} - {p_{12}}} \right)(1 - \cos 4\theta )} \right]}\\
	\begin{array}{c}
		{{p}_{16}^\prime} = {{p}_{61}^\prime} = \frac{1}{4}\sin 4\theta \left( {{p_{11}} - {p_{12}} - 2{p_{44}}} \right)\\
		{{p}_{26}^\prime} = {{p}_{62}^\prime} =  - {{p}_{16}^\prime}.
	\end{array}
\end{array}
\end{equation}

The integral in the photoelastic coupling term is expressed as
	\begin{equation}
	\int { {\epsilon _0}{n^4}E_i^*{E_j}{p_{ijkl}}{S_{kl}}{\mathrm{~d}}V}  = \int {{\epsilon _0}{n^4}(p_{11}' {f_{11}} + p_{12}' {f_{12}} + p_{13}' {f_{13}} + p_{31}' {f_{31}} + p_{33}' {f_{33}} + p_{44}' {f_{44}} + p_{66}' {f_{66}} + p_{16}' {f_{16}})} \mathrm{~d}V,
\end{equation}
where
\begin{equation}
	\begin{array}{c}{f_{11}=\left|E_{x}\right|^{2} S_{x x}+\left|E_{y}\right|^{2} S_{y y}} \\ {f_{12}=\left|E_{x}\right|^{2} S_{y y}+\left|E_{y}\right|^{2} S_{x x}} \\ {f_{13}=\left|E_{x}\right|^{2} S_{z z}+\left|E_{y}\right|^{2} S_{z z}} \\ {f_{31}=\left|E_{z}\right|^{2} S_{x x}+\left|E_{z}\right|^{2} S_{y y}} \\ {f_{33}=\left|E_{z}\right|^{2} S_{z z}} \\ {f_{44}=2 \mathrm{Re}\left\{E_{y}^{*} E_{z}\right\} \cdot 2 S_{y z}+2 \mathrm{Re}\left\{E_{x}^{*} E_{z}\right\} \cdot 2 S_{x z}} \\ {f_{16}=\left(\left|E_{x}\right|^{2}-\left|E_{y}\right|^{2}\right) \cdot 2 S_{x y}+2 \mathrm{Re}\left\{E_{x}^{*} E_{y}\right\} \cdot\left(S_{x x}-S_{y y}\right)}.
	\end{array}
\end{equation}

When $\theta = 0\degree$ and $45\degree$, there are mirror planes $\sigma_x$ and $\sigma_y$. We list the symmetry of $E_i^*E_j$, $S_{kl}$ and $f_{ij}$ in Tables \ref{ESsx} and \ref{Fsx}, where ‘e' stands for even and ‘o' stands for odd.
	\begin{table}[htbp]
	\centering
	\renewcommand\arraystretch{1.3}
	\begin{tabular}{|c|c|cccccc|}
		\hline
		&       & \multicolumn{1}{c}{$\left|E_{x}\right|^{2}$} & \multicolumn{1}{c}{$\left|E_{y}\right|^{2}$} & \multicolumn{1}{c}{$\left|E_{z}\right|^{2}$} & \multicolumn{1}{c}{${\mathop{\rm Re}\nolimits} \{ E_x^*{E_y}\}$} & \multicolumn{1}{c}{${\mathop{\rm Re}\nolimits} \{ E_x^*{E_z}\}$} & ${\mathop{\rm Re}\nolimits} \{ E_y^*{E_z}\}$ \bigstrut[b]\\
		\hline
		$\chi ({\sigma _x}) =  \pm 1$ & under $\sigma_x$ & e     & e     & e     & o     & o     & e \bigstrut\\
		\hline
		$\chi ({\sigma _y}) =  \pm 1$& under $\sigma_y$ & e     & e     & e     & o     & e     & o \bigstrut\\
		\hline
		&       & \multicolumn{1}{c}{$S_{xx}$} & \multicolumn{1}{c}{$S_{yy}$} & \multicolumn{1}{c}{$S_{zz}$} & \multicolumn{1}{c}{$S_{xy}$} & \multicolumn{1}{c}{$S_{xz}$} & $S_{yz}$ \bigstrut\\
		\hline
		$\chi ({\sigma _x}) =  1$   & under $\sigma_x$ & e     & e     & e     & o     & o     & e \bigstrut\\
		\hline
		$\chi ({\sigma _x}) =  -1$    & under $\sigma_x$ & o     & o     & o     & e     & e     & o \bigstrut\\
		\hline
		$\chi ({\sigma _y}) =  1$     & under $\sigma_y$ & e     & e     & e     & o     & e     & o \bigstrut\\
		\hline
		$\chi ({\sigma _y}) =  -1$    & under $\sigma_y$ & o     & o     & o     & e     & o     & e \bigstrut\\
		\hline
	\end{tabular}%
	\label{ESsx}
	\caption{The behavior of $E_{i}^*E_{j}$ and $S_{ij}$ under $\sigma_x$ and $\sigma_y$.}
\end{table}
\begin{table}[htbp]
	\renewcommand\arraystretch{1.3}
	\centering
	\begin{tabular}{|c|c|cccccccc|}
		\hline
		&       & $f_{11}$   & $f_{12}$   & $f_{13}$   & $f_{31}$   & $f_{33}$   & $f_{44}$   & $f_{66}$   & $f_{16}$ \bigstrut\\
		\hline
		$\chi ({\sigma _x}) =  1$     & under $\sigma_x$ & e     & e     & e     & e     & e     & e     & e     & o \bigstrut\\
		\hline
		$\chi ({\sigma _x}) =  -1$    & under $\sigma_x$ & o     & o     & o     & o     & o     & o     & o     & e \bigstrut\\
		\hline
		$\chi ({\sigma _y}) =  1$     & under $\sigma_y$ & e     & e     & e     & e     & e     & e     & e     & o \bigstrut\\
		\hline
		$\chi ({\sigma _y}) =  -1$    & under $\sigma_y$ & o     & o     & o     & o     & o     & o     & o     & e \bigstrut\\
		\hline
	\end{tabular}%
	\label{Fsx}%
	\caption{The behavior of $f_{ij}$ under $\sigma_x$ and $\sigma_y$.}
\end{table}%
We find that if the mechanical mode is odd under either $\sigma_x$ or $\sigma_y$, all the $f_{ij}$s except $f_{16}$ are odd and their integrals are zero. For $f_{16}$, because $p_{16}'=0$ when $\theta=0\degree$ or $45\degree$, the term $\int {p_{16}'{f_{16}}\mathrm{~d}V}$ also vanishes. For $\theta = 15\degree$ and $30\degree$, $g_{\mathrm{OM}, \mathrm{PE}}$ is in general nonzero.

In summary, $A_2$ modes for $\theta=0\degree$ and $45\degree$, which are BICs, have a zero optomechanical coupling. For other cases, optomechanical coupling is in general nonzero. This is also summarized in Table I of the main text.

\section{Optomechanical coupling of finite optomechanical crystals}
The optomechanical coupling for a sufficiently large optomechanical crystal can be calculated with the result of the unit-cell coupling. Performing the integrals of Eqs. \ref{gmb} and \ref{gpe} in the whole optomechanical crystal and apply the Bloch theorem, i.e., $\bm{Q}(\bm{r} + \bm{R}) = e^{i{\bm{k}_m} \cdot \bm{R}}\bm{Q}(\bm{r})$ and $\bm{E}(\bm{r} + \bm{R}) = e^{i{\bm{k}_o} \cdot \bm{R}}\bm{E}(\bm{r})$, where $\bm{k}_m$ and $\bm{k}_o$ are the Bloch wavevector of the mechanical and optical modes, respectively, we have
\begin{equation}
	{\bar g_{\mathrm{OM,MB}}} =  - \frac{{{\omega _o}}}{2}\frac{{\sum\limits_{\bm{R}} {{e^{i{\bm{k}_m} \cdot \bm{R}}}} \int_{\mathrm{u.c.}} {(\bm{Q} \cdot \hat{\bm{n}})(\Delta \epsilon {{\left| {{\bm{E}_\parallel }} \right|}^2} - \Delta {\epsilon ^{ - 1}}{{\left| {{\bm{D}_\bot }} \right|}^2})\mathrm{~d} S} }}{{\sum\limits_{\bm{R}} {\int_{\mathrm{u.c.}} {{\bm{E}^ * } \cdot \bm{D}\mathrm{~d} V} } }} = \frac{{\sum\limits_{\bm{R}} {{e^{i{\bm{k}_m} \cdot \bm{R}}}} }}{N_c}{g_{\mathrm{OM,MB}}}
\end{equation}
where $N_c$ is the number of unit cells and $\sum\limits_{\bm{R}}$ is the summation for all the lattice vectors. Same expression holds for $g_{\mathrm{OM,PE}}$. Because $\sum\limits_{\bm{R}} {{e^{i{\bm{k}_m} \cdot \bm{R}}}}$ is nonzero only when $\bm{k}_m=0$, optomechanical coupling exists only for mechanical modes at the $\Gamma$ point.
Because the total mass ${\bar m_{\mathrm{eff}}}=N_c{m_{\mathrm{eff}}}$, the optomechanical coupling of a sufficiently large optomechanical crystal is related to its unit-cell coupling as
\begin{equation}
	\bar g = \sqrt {\frac{\hbar }{{2N_c{m_{\mathrm{eff}}}{\omega _m}}}} ({g_{\mathrm{OM,MB}}} + {g_{\mathrm{OM,PE}}}) = \frac{{{g}}}{{\sqrt N_c }}.
\end{equation}

For finite optomechanical crystals, we have to consider the field envelop of the standing-wave resonances. The fields across unit cells now are related by $\bm{Q}(\bm{r} + \bm{R}) = e^{i{\bm{k}_m} \cdot \bm{R}}{\phi _{({p_m},{q_m})}}(\bm{R})\bm{Q}(\bm{r})$ and $\bm{E}(\bm{r} + \bm{R}) = e^{i{\bm{k}_o} \cdot \bm{R}}{\phi _{({p_o},{q_o})}}(\bm{R}) \bm{E}(\bm{r})$, where ${\phi _{({p},{q})}}(\bm{R})$ is the envelop function of the $(p,q)$-th order standing-wave resonance. The envelope function can be approximately solved as the eigenfunction of a flat-top potential well within the boundary of the finite optomechanical crystal. Then we have
\begin{equation}
	\begin{aligned}
		{\bar g_{\mathrm{OM,MB}}} &=  - \frac{{{\omega _o}}}{2}\frac{{\sum\limits_{\bm{R}} e^{i{\bm{k}_m} \cdot \bm{R}}{{\phi _{({p_m},{q_m})}}(\bm{R}){{\left| {{\phi _{({p_o},{q_o})}}(\bm{R})} \right|}^2}} \int_{\mathrm{u.c.}} {(\bm{Q} \cdot \hat{ \bm{n}})(\Delta \epsilon {{\left| {{\bm{E}_\parallel }} \right|}^2} - \Delta {\epsilon^{ - 1}}{{\left| {{\bm{D}_ \bot }} \right|}^2})\mathrm{~d}S} }}{{\sum\limits_{\bm{R}} {{{\left| {{\phi _{(p,q)}}(\bm{R})} \right|}^2}\int_{\mathrm{u.c.}} {{\bm{E}^ * } \cdot \bm{D}\mathrm{~d}V} } }}\\
		&= \frac{{\sum\limits_{\bm{R}} e^{i{\bm{k}_m} \cdot \bm{R}}{{\phi _{({p_m},{q_m})}}(\bm{R}){{\left| {{\phi _{({p_o},{q_o})}}(\bm{R})} \right|}^2}} }}{{\sum\limits_{\bm{R}} {{{\left| {{\phi _{(p_o,q_o)}}(\bm{R})} \right|}^2}} }}{g_{\mathrm{OM,MB}}}\\
		&\approx \frac{{\int e^{i{\bm{k}_m} \cdot \bm{R}}{{\phi _{(p_m,q_m)}}(\bm{R}){{\left| {{\phi _{(p_o,q_o)}}(\bm{R})} \right|}^2}\mathrm{~d}S} }}{{\int {{{\left| {{\phi _{(p_o,q_o)}}(\bm{R})} \right|}^2}\mathrm{~d}S} }}{g_{\mathrm{OM,MB}}}.
	\end{aligned}
\end{equation}
where the last integrals are performed in the whole optomechanical crystal. Similar expression applies to the photoelastic term. For the total effective mass, we have
\begin{equation}
	\begin{aligned}
			{\bar m_{{\mathrm{eff}}}} =& \sum\limits_{\bm{R}} {\int_{\mathrm{u.c.}} {\rho {{\left| \bm{Q} \right|}^2}\mathrm{~d}V} }  = \sum\limits_{\bm{R}} {{{\left| {{\phi _{(p_m,q_m)}}(\bm{R})} \right|}^2}\int_{\mathrm{u.c.}} {\rho {{\left| \bm{Q} \right|}^2}\mathrm{~d}V} }  = \sum\limits_{\bm{R}} {{{\left| {{\phi _{(p_m,q_m)}}(\bm{R})} \right|}^2}} {m_{\mathrm{eff}}}\approx N_c\frac{{\int {{{\left| {{\phi _{(p_m,q_m)}}(\bm{R})} \right|}^2}\mathrm{~d}S} }}{{\int {\mathrm{d}S} }}m_{\mathrm{eff}}.
	\end{aligned}
\end{equation}
Finally, the optomechanical coupling of a finite optomechanical crystal is related to its unit-cell coupling by
\begin{equation}
	\bar g = \frac{{\zeta {g}}}{{\sqrt N_c }},
\end{equation}
where the finite-size correction factor $\zeta$ is given by
\begin{equation}
	\zeta  = \frac{{\int e^{i{\bm{k}_m} \cdot \bm{R}}{{\phi _{(p_m,q_m)}}(\bm{R}){{\left| {{\phi _{(p_o,q_o)}}(\bm{R})} \right|}^2}\mathrm{~d}S} }}{{\int {{{\left| {{\phi _{(p_o,q_o)}}(\bm{R})} \right|}^2}\mathrm{~d}S\sqrt {\int {{{\left| {{\phi _{(p_m,q_m)}}(\bm{R})} \right|}^2}\mathrm{~d}S} /\int {\mathrm{d}S} } } }}.
\end{equation}
For the mechanical Bloch wavevector $\bm{k}_m=0$, the numerically-calculated correction factor for a few standing-wave mechanical and optical resonances of a hexagonal crystal is listed in Table \ref{tab:addlabel}.  The field envelops of some orders are shown in Fig. 3c. The order of the hexagonal crystal mode can be hard to identify. We determined it by adiabatically deforming the hexagonal crystal to a rectangular crystal and tracing the corresponding mode, where $(p, q)$-th resonance has $p-1$ and $q-1$ nodes along the $x$ and $y$ direction, respectively.

\begin{table}[htbp]
	\centering
	\caption{Correction factor $\zeta$ for some standing-wave mechanical and optical resonances of a hexagonal crystal.}
	\begin{tabular}{|cc|ccccc|}
		\hline
		& Optical & \multirow{2}[2]{*}{(1,1)} & \multirow{2}[2]{*}{(1,2)} & \multirow{2}[2]{*}{(2,1)} & \multirow{2}[2]{*}{(2,2)} & \multirow{2}[2]{*}{(3,1)} \bigstrut[t]\\
		Mechanical &       &       &       &       &       &  \bigstrut[b]\\
		\hline
		\multicolumn{2}{|c|}{(1,1)} & 1.4075 & 1.1427 & 1.1415 & 0.9737 & 0.9740 \bigstrut[t]\\
		\multicolumn{2}{|c|}{(3,1)} & 0.0185     & 0.8250     & 0.8322     & 0.0183     & 0.0805 \\
		\multicolumn{2}{|c|}{(1,3)} & 0.3987     & 0.4442     & 0.3692     & 0.7307     & 0.7398 \bigstrut[b]\\
		\hline
	\end{tabular}%
	\label{tab:addlabel}%
\end{table}%

The interaction Hamiltonian of the finite optomechanical crystal involving a pair of mechanical and optical standing-wave resonances is given by $\hat H=\bar g\hat a^\dagger\hat a( \hat b^\dagger+\hat b)$, where $\hat a^\dagger$($\hat a$) and $\hat b^\dagger$($\hat b$) are the creation(annihilation) operators of the optical and mechanical resonances, respectively. When the optomechanical crystal is driven by a pump detuned from the optical resonance by the mechanical frequency, the interaction Hamiltonian could be linearized to be $\hat H=G(\hat a^\dagger \hat b^\dagger+\hat a\hat b)$ for blue-detuned pump or $\hat H=G(\hat a^\dagger \hat b+\hat a\hat b^\dagger)$ for red-detuned pump, with 
\be
G=\bar g\sqrt{n}=\zeta g\sqrt{n_c}
\ee
the parametrically-enhanced optomechanical coupling, where $n$ and $n_c=n/N_c$ are the total photon number in the optical resonance and average photon number per unit cell, respectively.

\section{Mechanical noise power spectrum}
The total noise power spectral density $S$ measured by the photodetector is given by \cite{meenehan2014silicon,safavi2013laser}
\be\label{spectrum}
\begin{aligned}
	 S &= {S_e}+\frac{G_e^2}{R_I}\left(  {S_{\mathrm{EDFA}}} + G^2_{\mathrm{EDFA}}{S^2_{\mathrm{SN}}}{\Big(1 + \eta \frac{{{\kappa _e}}}{\kappa }\frac{{8{G^2}}}{\kappa }{{S}_m}[\omega]\Big)} \right),
\end{aligned}
\ee
where $S_e$ is the electronic noise of the detector, $S_{\mathrm{EDFA}}$ is the noise of EDFA, $S_{\mathrm{SN}}=\sqrt{P_{\mathrm{out}}\hbar\omega_l}$ is the optical shot noise,, $S_m[\omega]$ is the mechanical noise spectrum, $\kappa=\kappa_i+\kappa_e$ is the total optical dissipation, $G$ is the parametrically-enhanced optomechanical coupling, $\eta$ is the total detection efficiency, $G_e$ is the detector gain factor from optical power to voltage, $G_{\mathrm{EDFA}}$ is the EDFA gain, and $R_I$ is the input impedance of the spectrum analyzer. The optically-transduced mechanical noise spectrum is given by
\begin{equation}\label{mechspec}
	{S_m}[\omega] = \frac{1}{2}\left(\frac{{\gamma (\bar n_m + 1)}}{{{{\left( {{\omega _m} - \omega } \right)}^2} + {{\left( {\gamma /2} \right)}^2}}}+\frac{{\gamma \bar n_m}}{{{{\left( {{\omega _m} +\omega } \right)}^2} + {{\left( {\gamma /2} \right)}^2}}}\right)
\end{equation}
where $\bar n_m=\frac{k_BT}{\hbar\omega_m}$ the thermal occupation of the mechanical mode. The total detection efficiency is $\eta=\eta_{\mathrm{cpl}}\eta_{\mathrm{t}}\eta_{\mathrm{det}}$, where $\eta_\mathrm{cpl}$ is the coupling efficiency of the grating coupler at the pump wavelength, $\eta_{\mathrm{t}}$ is the total transmission efficiency in the optical fiber path, and $\eta_{\mathrm{det}}$ is the quantum efficiency of the photodetector. We measured $\eta_\mathrm{t}=0.80$ and $\eta_\mathrm{cpl}\approx 0.5$ depending on the pump wavelength, while $\eta_\mathrm{det}=0.68$ is given by the detector, which gives $\eta\approx 0.27$.

We normalize the noise power spectrum with respect to the total noise floor which yields 
\be\label{nspectrum}
\begin{aligned}
	\overline S &= 1 + \frac{{S'_{\mathrm{SN}}}}{{S_e}+ {S'_{\mathrm{EDFA}}} + {S'_{\mathrm{SN}}}} \eta \frac{{{\kappa _e}}}{\kappa }\frac{{8{G^2}}}{\kappa }{{S}_m}[\omega],
\end{aligned}
\ee
where we have used $ {S'_{\mathrm{EDFA}}} \equiv \frac{G_e^2}{R_I} {S_{\mathrm{EDFA}}} $ and $S'_{\mathrm{SN}}\equiv \frac{G_e^2G_{\mathrm{EDFA}}}{R_I}{S_{\mathrm{SN}}}$ to denote the photodetector-measured EDFA noise and laser shot noise. The electronic noise $S_e$ is measured by blocking all the light. $S'_{\mathrm{SN}}+S_e$ can be determined by removing the EDFA while keeping the laser power incident to the detector unchanged. Thus we obtain $S'_{\mathrm{SN}}$ and the ratio of $\frac{{S'_{\mathrm{SN}}}}{{S_e}+ {S'_{\mathrm{EDFA}}} + {S'_{\mathrm{SN}}}}$. Finally, the measured normalized noise power spectrum is fitted using Eqs. \ref{nspectrum} and \ref{mechspec}, with $\omega_m$, $\gamma$, and $G$ the only fitting parameters. The unit-cell optomechanical coupling is calculated from $g=G/(\zeta \sqrt{n_c})$, using $n_c=\kappa_eP/((\omega_m)^2+(\kappa/2)^2)/\frac{3\sqrt{3}}{2}N^2$ and $\zeta$ depending on the optical standing-wave resonance that is deployed (see Table \ref{tab:addlabel}). For the four devices shown in Fig. 4, we used optical resonances of the following order, $0\degree$ (1,1), $15\degree$ (1,1), $30\degree$ (1,1), $45\degree$ (2,1), while the observed mechanical resonances are (1,1).

\section{Fabrication results}
The silicon-on-insulator microchip (220 nm silicon device layer and 3 $\upmu$m buried oxide layer) is first rinsed with Acetone, isopropyl alcohol and deionized water to clean the surface. The pattern is then defined using electron beam lithography (Elionix ELS-G150) with 300 nm thick ZEP520A mask, followed by inductively coupled plasma reactive ion etch (Oxford PlasmaPro 100 Cobra) of silicon using SF6 and CHF3 . The mask residual is removed by n-methyl-2-pyrrolidone hot bath and oxygen plasma cleaning (Diener Descum). 

\begin{figure*}[htp]
	\centering
	\includegraphics[width=0.7\linewidth]{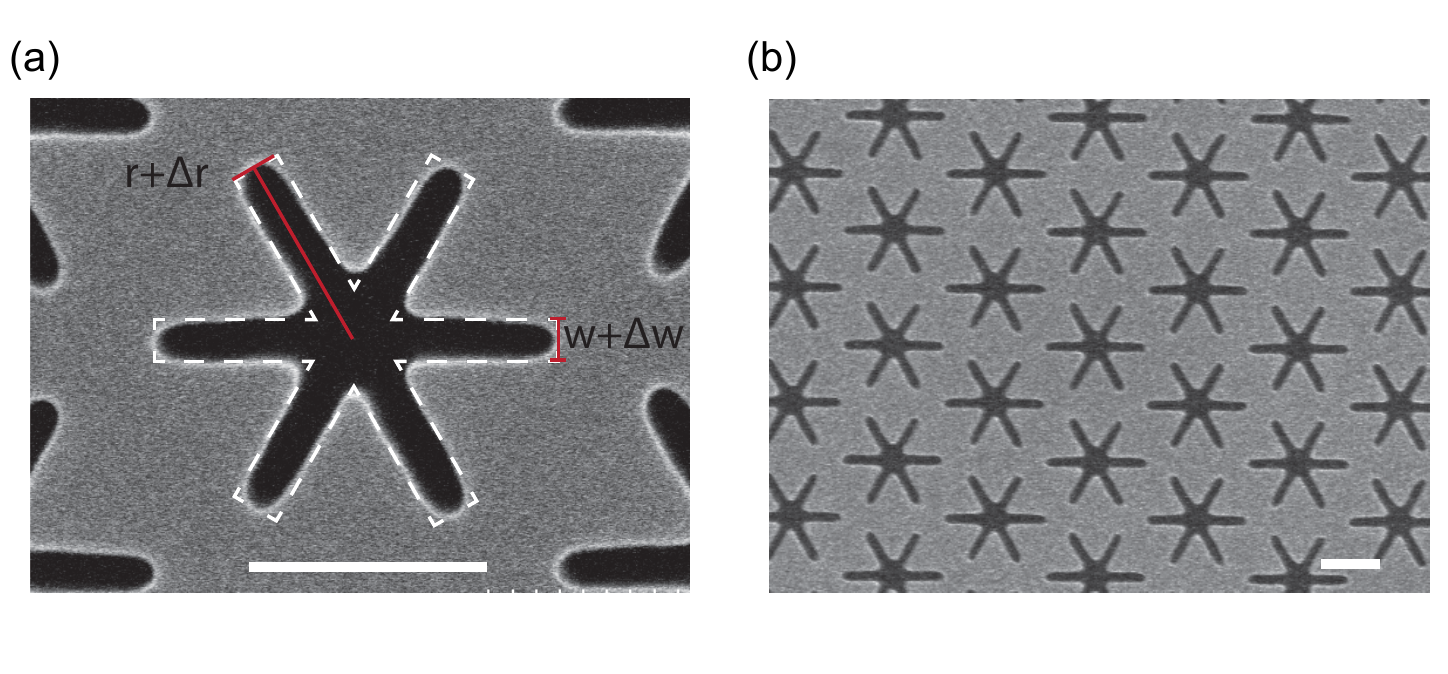}
	\caption{\textbf{a}. An SEM image of a typical unit cell. The dashed line shows the fitted snowflake structure. \textbf{b}. An SEM image of a few unit cells within an optomechanical crystal. The white bar refers to 200 nm for both images.}
	\label{fig:figs5}
\end{figure*}

The SEM image of a typical unit cell is shown in Fig.\ref{fig:figs5}a. The disorder of the snowflake patterns are measured both inside a same optomechanical crystal and across different optomechanical crystals (Table \ref{tab:distortion}). The standard deviation of $r$ and $w$ is about 2.7\% and 7.22\% for the former and 2.6\% and 6.7\% for the latter. For comparison of different optomechanical crystals, we examined unit cells from the same location. Roughly speaking, the variation across different devices accounts for the the resonance frequency fluctuation and disorder within an optomechanical crystal induces scattering loss and decrease the resonance quality factor (see Section \ref{sim}).
\begin{table}[htbp]
	\centering
	\caption{Mean and standard deviation of $r$ and $w$.}
	\begin{tabular}{|c|c|c|}
		\hline
		 & Across different devices & Inside one device\\
		\hline
		$\bar{r}$ (nm)  &   172.09   &  167.35    \\
		\hline
		$r_{\mathrm{max}}$ (nm) & 177.60 & 174.65\\
		\hline
		$r_{\mathrm{min}}$ (nm) & 161.00 & 154.65\\
		\hline
		$\Delta r$ (nm) & 4.49& 4.57\\
		\hline
		$\bar{w}$ (nm) &   37.76    &  37.26    \\
		\hline
		$w_{\mathrm{max}}$ (nm) & 44.00& 43.10\\
		\hline
		$w_{\mathrm{min}}$ (nm) &31.70 & 32.70\\
		\hline
		$\Delta w$ (nm)& 2.53& 2.67\\
		\hline
	\end{tabular}
	\label{tab:distortion}
\end{table}

\section{Simulation of Scattering Loss}\label{sim}

\begin{figure*}[htp]
	\centering
	\includegraphics[width=0.7\linewidth]{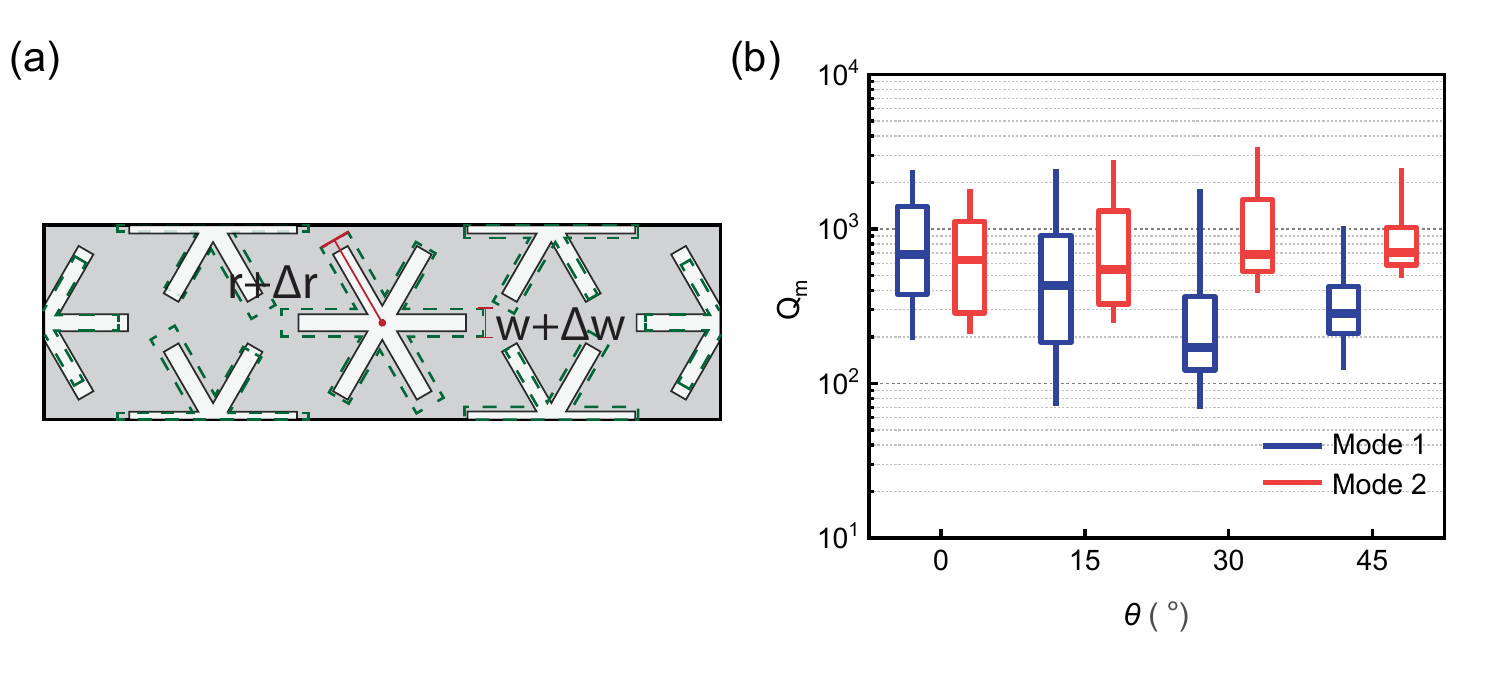}
	\caption{\textbf{a}. A schematic diagram of a $1 \times 4$ super-cell structure with disorders. \textbf{b}. The simulated $Q_m$ for different $\theta$. The lower and upper edges of the boxes represent the 16\% and 84\% bounds of the data, respectively, and the middle lines is the median value.}
	\label{fig:figs4}
\end{figure*}

In multimode photonic/phononic crystals, structural disorders could cause scattering among different modes and thus introduce more loss channels to a given mode \cite{jin2019topologically}. To estimate the quality factor of standing-wave mechanical resonances in disordered crystals, we simulated a $1\times 4$ super-cell structure (Fig. \ref{fig:figs4}a),
where variations of both length ($\Delta r$) and width ($\Delta w$) of each snowflake hole are introduced. Each $\Delta r$ and $\Delta w$ are independently sampled and the standard deviation of $\Delta r$ and $\Delta w$ are controlled to be 4 nm and 2 nm, respectively, in accordance with the fabricated devices. We run 50 sets of simulations each for $\theta = 0\degree$, $15\degree$, $30\degree$, and $45\degree$. We set the Bloch wavevector to be $k=2\pi/(3aN)$ with $N$ = 25, approximately corresponding to the $(1, 1)$  standing-wave resonance in size $N=25$ hexagonal optomechanical crystals. The simulated radiative quality factors $Q_r$ combined with room-temperature material absorption $Q_a$, which we assumed to be 3000, gives the total mechanical quality factor: $1/Q_m=1/Q_r+1/Q_a$.  The result is plotted in Fig. \ref{fig:figs4}b. Due to the small size of the optomechanical crystal, there are no significant variations of $Q_m$ among four different $\theta$'s. We use the data set of all four angles here to calculate the upper and lower bounds of $Q_m$ as shown in Fig. 4.

\section{Design of grating coupler}
\begin{figure*}[h]
	\centering
	\includegraphics[width=1\linewidth]{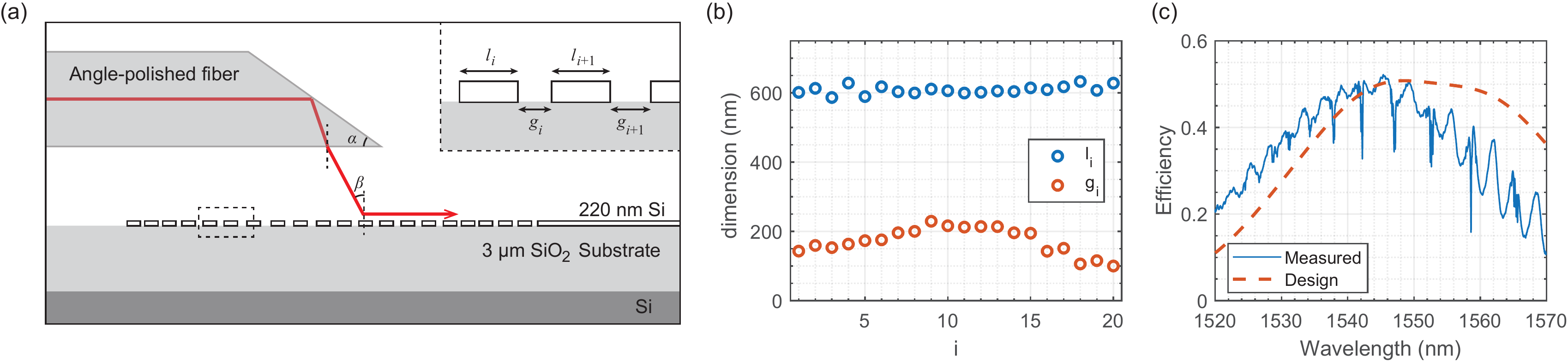}
	\caption{\textbf{a}. Schematic plot of the angle-polished fiber and grating coupler. \textbf{b}. Parameters $l_i$ and $g_i$ for the optimized grating coupler. \textbf{c}. Measured and simulated efficiency of the optimized grating coupler.}
	\label{fig:gratingcouplerspectrum}
\end{figure*}
Fig. \ref{fig:gratingcouplerspectrum}a shows the fiber-optic coupler for the measurement of the optomechanical crystal. Light is guided by an angle-polished single-mode fiber, reflects at the polished surface, and is incident onto a grating coupler. The angle $\alpha$ of the fiber is chosen to ensure total internal reflection. Because of the inevitable air gap between the fiber and grating coupler, the incident angle onto the grating coupler is given by $\beta = \arcsin(n_{\mathrm{fiber}}\cdot\sin(90\degree - 2\alpha))$.
In general, the grating coupler is designed to convert the incident light to a guided wave by satisfying the phase-matching condition $	\frac{{2\pi {n_{\mathrm{eff}}}}}{\lambda } = \frac{{2\pi }}{\lambda }\sin \beta  + \frac{{2\pi }}{\Lambda }$,
where $n_{\mathrm{eff}}$ is the effective refractive index of the grating and $\Lambda$ is the lattice constant of the grating. It turns out an apodized grating coupler could yield higher coupling efficiency than a uniform grating coupler because of the better mode-matching with the single-mode optical fiber \cite{zhao2020design}. Thus, we adopt the apodized grating coupler scheme.   The apodized grating coupler consists of 20 unit-cells with 40 individual tunable parameters $l_i$ and $g_i$. The parameters are optimized using a standard inverse design approach \cite{molesky2018inverse} to maximize the coupling efficiency for different fiber angles. The optimized fiber angle is found to be 35.5$^\degree$ and the corresponding $l_i$ and $g_i$ are shown in Fig. \ref{fig:gratingcouplerspectrum}b. Fig. \ref{fig:gratingcouplerspectrum}c shows the simulated transmission and measured device spectrum (inferred from the reflection spectrum). The peak efficiency is observed to be $52$\% with a 3-dB bandwidth of $40\,$nm.

\section{Design of photonic crystal mirror}

\begin{figure}[htp]
	\centering
	\includegraphics[width=1\linewidth]{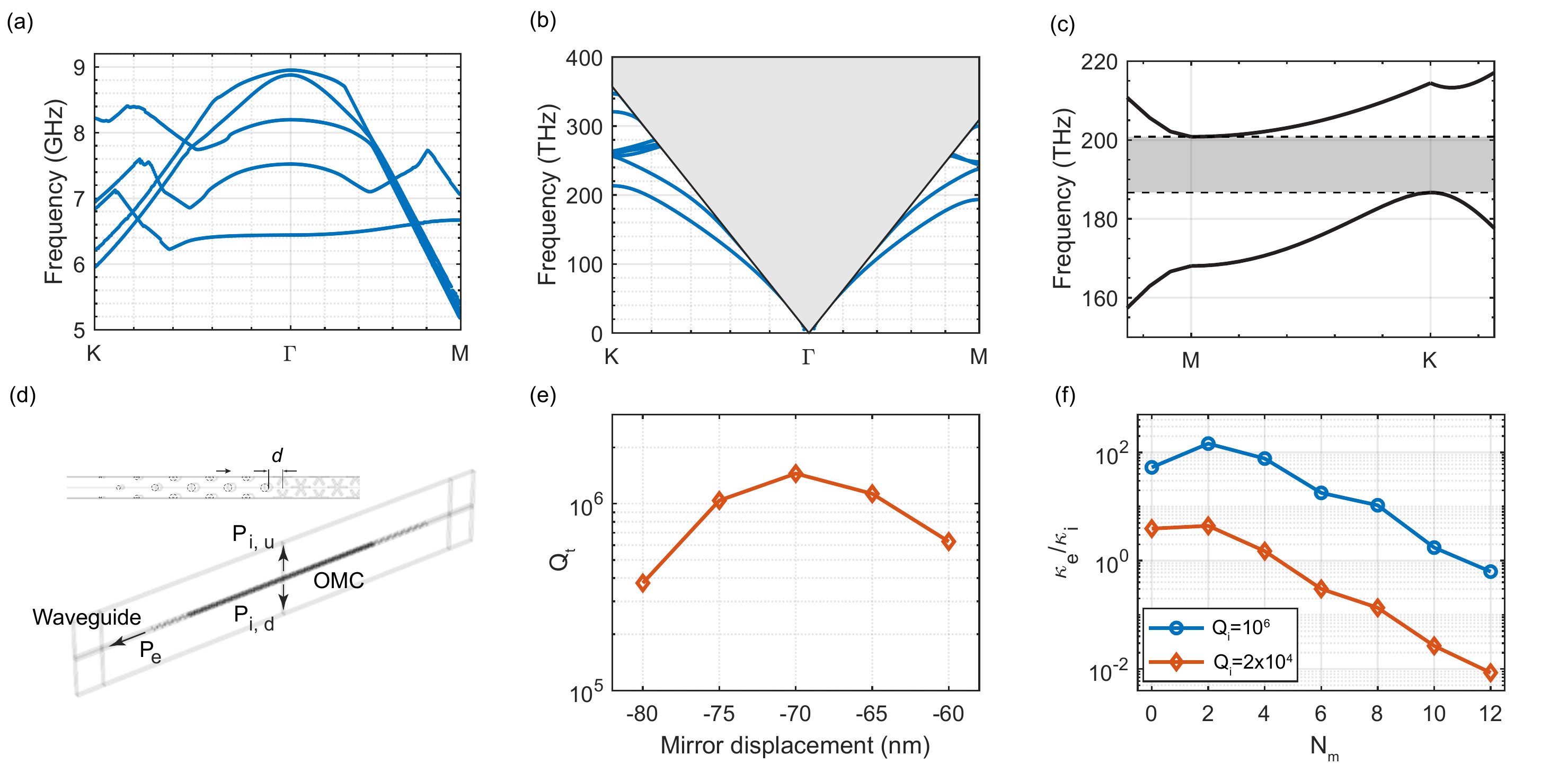}
	\caption{\textbf{a} and \textbf{b} Mechanical and optical bandstructure of the nominal optomechanical crystal.   \textbf{c}. Band structure of the photonic crystal mirror. \textbf{d}. Super-cell used in the simulation. Inset: Top view of the junction between the optomechanical crystal and waveguide. Dashed circles are the original location of the photonic crystal mirror holes before displacement. \textbf{e}. Simulated total quality factor $Q_t$ versus the displacement $d-a$. \textbf{f}. Simulated $\kappa_e/\kappa_i$ versus the number of nominal mirror layers $N_m$ for $Q_i=10^6$ and $Q_i=2\times 10^4$. }
	\label{fig:figs3v1}
\end{figure}

Fig. \ref{fig:figs3v1}a and b show the full mechanical and optical bandstructure of the nominal optomechanical crystal. 
The photonic crystal mirror consists of a triangular lattice of cylindrical holes with $r=78\,$nm and the same lattice constant $a = 389\,$nm as the snowflake optomechanical crystal. The simulated band structure with the two lowest TE-like bands
 (Fig. \ref{fig:figs3v1}c) shows a complete bandgap from $187\,$THz ($1606\,$nm) to $201\,$THz ($1493\,$nm). The photonic crystal mirror effectively suppresses the side-leakage but also perturbs the tail of the standing-wave resonance, inducing wavevector components that fall inside the continuum and causing out-of-plane radiation. We could alleviate such out-of-plane radiation by tuning the separation $d$ between the photonic crystal mirror and optomechanical crystal \cite{akahane2005fine,chen2021observation}. To simulate this, we take a strip of optomechanical crystal of 25 unit cells as one super-cell and add periodic boundary conditions on the transverse direction, as shown in Fig. \ref{fig:figs3v1}d. A sufficient number of photonic crystal mirror unit-cells are included and so the total quality factor $Q_t$ of the resonance is dominated by the out-of-plane radiation. Fig. \ref{fig:figs3v1}e shows $Q_t\equiv\frac{\mathrm{Re}[\omega_o]}{\mathrm{Im}[\omega_o]}$ of the fundamental standing-wave resonance for different $d$. An optimal $d=319$ nm is observed, corresponding to a displacement of the photonic crystal from the nominal lattice constant by $d-a=-70$ nm. 

The fabricated optomechanical crystal connects to a waveguide and the external coupling rate could be controlled by the number of phtonic crystal mirror layers between the waveguide and optomechanical crystal. In the simulation, we change the number of phtonic crystal mirror layers $N_m$ on one side (the waveguide side) of the optomechanical crystal super-cell while keeping the other side with a sufficient amount of photonic crystal mirror layers. We have also included four additional tapered mirror layers on the waveguide side. The external and internal quality factors are inferred from the energy flux as
\begin{equation}
	{Q_e} = \frac{{{P_e} + {P_{i,u}} + {P_{i,d}}}}{{{P_e}}}{Q_t},\quad {Q_i} = \frac{{{P_e} + {P_{i,u}} + {P_{i,d}}}}{{{P_{i,u}} + {P_{i,d}}}}{Q_t},
\end{equation}
where $P_e$ is the energy flux towards the waveguide and $P_{i,u}$ and $P_{i,d}$ are out-of-plane energy flux upward and downward, respectively. Fig. \ref{fig:figs3v1}f shows $\kappa_e/\kappa_i\equiv Q_i/Q_e$. The simulation is done for two cases. One case is with a large number of photonic crystal mirrors on the other side of the super-cell such that $Q_i\approx 10^6$. In the other case, we purposely reduced the number of photonic crystal mirrors on the other side of the super-cell such that $Q_i\approx 2\times10^4$ to mimic the fabricated optomechanical crystal. For this case. We find the critical coupling $\kappa_e/\kappa_i\approx1$ is achieved around $N_m=4$.


\begin{thebibliography}{10}
\expandafter\ifx\csname url\endcsname\relax
  \def\url#1{\texttt{#1}}\fi
\expandafter\ifx\csname urlprefix\endcsname\relax\def\urlprefix{URL }\fi
\providecommand{\bibinfo}[2]{#2}
\providecommand{\eprint}[2][]{\url{#2}}

\bibitem{aspelmeyer2014cavity}
\bibinfo{author}{Aspelmeyer, M.}, \bibinfo{author}{Kippenberg, T.~J.} \&
  \bibinfo{author}{Marquardt, F.}
\newblock \bibinfo{title}{Cavity optomechanics}.
\newblock \emph{\bibinfo{journal}{Reviews of Modern Physics}}
  \textbf{\bibinfo{volume}{86}}, \bibinfo{pages}{1391} (\bibinfo{year}{2014}).

\bibitem{eichenfield2009optomechanical}
\bibinfo{author}{Eichenfield, M.}, \bibinfo{author}{Chan, J.},
  \bibinfo{author}{Camacho, R.~M.}, \bibinfo{author}{Vahala, K.~J.} \&
  \bibinfo{author}{Painter, O.}
\newblock \bibinfo{title}{Optomechanical crystals}.
\newblock \emph{\bibinfo{journal}{Nature}} \textbf{\bibinfo{volume}{462}},
  \bibinfo{pages}{78--82} (\bibinfo{year}{2009}).

\bibitem{safavi2014two}
\bibinfo{author}{Safavi-Naeini, A.~H.}, \bibinfo{author}{Hill, J.~T.},
  \bibinfo{author}{Meenehan, S.}, \bibinfo{author}{Chan, J.},
  \bibinfo{author}{Gr{\"o}blacher, S.} \& \bibinfo{author}{Painter, O.}
\newblock \bibinfo{title}{Two-dimensional phononic-photonic band gap
  optomechanical crystal cavity}.
\newblock \emph{\bibinfo{journal}{Physical Review Letters}}
  \textbf{\bibinfo{volume}{112}}, \bibinfo{pages}{153603}
  (\bibinfo{year}{2014}).

\bibitem{ren2020two}
\bibinfo{author}{Ren, H.}, \bibinfo{author}{Matheny, M.~H.},
  \bibinfo{author}{MacCabe, G.~S.}, \bibinfo{author}{Luo, J.},
  \bibinfo{author}{Pfeifer, H.}, \bibinfo{author}{Mirhosseini, M.} \&
  \bibinfo{author}{Painter, O.}
\newblock \bibinfo{title}{Two-dimensional optomechanical crystal cavity with
  high quantum cooperativity}.
\newblock \emph{\bibinfo{journal}{Nature Communications}}
  \textbf{\bibinfo{volume}{11}}, \bibinfo{pages}{1--10} (\bibinfo{year}{2020}).

\bibitem{maccabe2020nano}
\bibinfo{author}{MacCabe, G.~S.}, \bibinfo{author}{Ren, H.},
  \bibinfo{author}{Luo, J.}, \bibinfo{author}{Cohen, J.~D.},
  \bibinfo{author}{Zhou, H.}, \bibinfo{author}{Sipahigil, A.},
  \bibinfo{author}{Mirhosseini, M.} \& \bibinfo{author}{Painter, O.}
\newblock \bibinfo{title}{Nano-acoustic resonator with ultralong phonon
  lifetime}.
\newblock \emph{\bibinfo{journal}{Science}} \textbf{\bibinfo{volume}{370}},
  \bibinfo{pages}{840--843} (\bibinfo{year}{2020}).

\bibitem{chan2011laser}
\bibinfo{author}{Chan, J.}, \bibinfo{author}{Alegre, T.~M.},
  \bibinfo{author}{Safavi-Naeini, A.~H.}, \bibinfo{author}{Hill, J.~T.},
  \bibinfo{author}{Krause, A.}, \bibinfo{author}{Gr{\"o}blacher, S.},
  \bibinfo{author}{Aspelmeyer, M.} \& \bibinfo{author}{Painter, O.}
\newblock \bibinfo{title}{Laser cooling of a nanomechanical oscillator into its
  quantum ground state}.
\newblock \emph{\bibinfo{journal}{Nature}} \textbf{\bibinfo{volume}{478}},
  \bibinfo{pages}{89--92} (\bibinfo{year}{2011}).

\bibitem{marinkovic2018optomechanical}
\bibinfo{author}{Marinkovi{\'c}, I.}, \bibinfo{author}{Wallucks, A.},
  \bibinfo{author}{Riedinger, R.}, \bibinfo{author}{Hong, S.},
  \bibinfo{author}{Aspelmeyer, M.} \& \bibinfo{author}{Gr{\"o}blacher, S.}
\newblock \bibinfo{title}{Optomechanical bell test}.
\newblock \emph{\bibinfo{journal}{Physical Review Letters}}
  \textbf{\bibinfo{volume}{121}}, \bibinfo{pages}{220404}
  (\bibinfo{year}{2018}).

\bibitem{wallucks2020quantum}
\bibinfo{author}{Wallucks, A.}, \bibinfo{author}{Marinkovi{\'c}, I.},
  \bibinfo{author}{Hensen, B.}, \bibinfo{author}{Stockill, R.} \&
  \bibinfo{author}{Gr{\"o}blacher, S.}
\newblock \bibinfo{title}{A quantum memory at telecom wavelengths}.
\newblock \emph{\bibinfo{journal}{Nature Physics}}
  \textbf{\bibinfo{volume}{16}}, \bibinfo{pages}{772--777}
  (\bibinfo{year}{2020}).

\bibitem{ludwig2013quantum}
\bibinfo{author}{Ludwig, M.} \& \bibinfo{author}{Marquardt, F.}
\newblock \bibinfo{title}{Quantum many-body dynamics in optomechanical arrays}.
\newblock \emph{\bibinfo{journal}{Physical Review Letters}}
  \textbf{\bibinfo{volume}{111}}, \bibinfo{pages}{073603}
  (\bibinfo{year}{2013}).

\bibitem{brendel2017pseudomagnetic}
\bibinfo{author}{Brendel, C.}, \bibinfo{author}{Peano, V.},
  \bibinfo{author}{Painter, O.~J.} \& \bibinfo{author}{Marquardt, F.}
\newblock \bibinfo{title}{Pseudomagnetic fields for sound at the nanoscale}.
\newblock \emph{\bibinfo{journal}{Proceedings of the National Academy of
  Sciences}} \textbf{\bibinfo{volume}{114}}, \bibinfo{pages}{E3390--E3395}
  (\bibinfo{year}{2017}).

\bibitem{brendel2018snowflake}
\bibinfo{author}{Brendel, C.}, \bibinfo{author}{Peano, V.},
  \bibinfo{author}{Painter, O.} \& \bibinfo{author}{Marquardt, F.}
\newblock \bibinfo{title}{Snowflake phononic topological insulator at the
  nanoscale}.
\newblock \emph{\bibinfo{journal}{Physical Review B}}
  \textbf{\bibinfo{volume}{97}}, \bibinfo{pages}{020102}
  (\bibinfo{year}{2018}).

\bibitem{ren2020topological}
\bibinfo{author}{Ren, H.}, \bibinfo{author}{Shah, T.},
  \bibinfo{author}{Pfeifer, H.}, \bibinfo{author}{Brendel, C.},
  \bibinfo{author}{Peano, V.}, \bibinfo{author}{Marquardt, F.} \&
  \bibinfo{author}{Painter, O.}
\newblock \bibinfo{title}{Topological phonon transport in an optomechanical
  system}.
\newblock \emph{\bibinfo{journal}{arXiv preprint arXiv:2009.06174}}
  (\bibinfo{year}{2020}).

\bibitem{sarabalis2017release}
\bibinfo{author}{Sarabalis, C.~J.}, \bibinfo{author}{Dahmani, Y.~D.},
  \bibinfo{author}{Patel, R.~N.}, \bibinfo{author}{Hill, J.~T.} \&
  \bibinfo{author}{Safavi-Naeini, A.~H.}
\newblock \bibinfo{title}{Release-free silicon-on-insulator cavity
  optomechanics}.
\newblock \emph{\bibinfo{journal}{Optica}} \textbf{\bibinfo{volume}{4}},
  \bibinfo{pages}{1147--1150} (\bibinfo{year}{2017}).

\bibitem{qi2021nonsuspended}
\bibinfo{author}{Qi, R.}, \bibinfo{author}{Xu, Q.}, \bibinfo{author}{Wu, N.},
  \bibinfo{author}{Cui, K.}, \bibinfo{author}{Zhang, W.} \&
  \bibinfo{author}{Huang, Y.}
\newblock \bibinfo{title}{Nonsuspended optomechanical crystal cavities using as
  2 s 3 chalcogenide glass}.
\newblock \emph{\bibinfo{journal}{Photonics Research}}
  \textbf{\bibinfo{volume}{9}}, \bibinfo{pages}{893--898}
  (\bibinfo{year}{2021}).

\bibitem{zhang2021silicon}
\bibinfo{author}{Zhang, J.}, \bibinfo{author}{Roux, X.~L.},
  \bibinfo{author}{Montesinos-Ballester, M.}, \bibinfo{author}{Ortiz, O.},
  \bibinfo{author}{Marris-Morini, D.}, \bibinfo{author}{Vivien, L.},
  \bibinfo{author}{Lanzillotti-Kimura, N.~D.} \& \bibinfo{author}{Alonso-Ramos,
  C.}
\newblock \bibinfo{title}{Silicon-on-insulator optomechanical microresonator
  with tight photon and phonon confinement}.
\newblock \emph{\bibinfo{journal}{arXiv preprint arXiv:2103.08465}}
  (\bibinfo{year}{2021}).

\bibitem{meenehan2014silicon}
\bibinfo{author}{Meenehan, S.~M.}, \bibinfo{author}{Cohen, J.~D.},
  \bibinfo{author}{Gr{\"o}blacher, S.}, \bibinfo{author}{Hill, J.~T.},
  \bibinfo{author}{Safavi-Naeini, A.~H.}, \bibinfo{author}{Aspelmeyer, M.} \&
  \bibinfo{author}{Painter, O.}
\newblock \bibinfo{title}{Silicon optomechanical crystal resonator at
  millikelvin temperatures}.
\newblock \emph{\bibinfo{journal}{Physical Review A}}
  \textbf{\bibinfo{volume}{90}}, \bibinfo{pages}{011803}
  (\bibinfo{year}{2014}).

\bibitem{tong2020observation}
\bibinfo{author}{Tong, H.}, \bibinfo{author}{Liu, S.}, \bibinfo{author}{Zhao,
  M.} \& \bibinfo{author}{Fang, K.}
\newblock \bibinfo{title}{Observation of phonon trapping in the continuum with
  topological charges}.
\newblock \emph{\bibinfo{journal}{Nature Communications}}
  \textbf{\bibinfo{volume}{11}}, \bibinfo{pages}{1--7} (\bibinfo{year}{2020}).

\bibitem{chen2016mechanical}
\bibinfo{author}{Chen, Y.}, \bibinfo{author}{Shen, Z.}, \bibinfo{author}{Xiong,
  X.}, \bibinfo{author}{Dong, C.-H.}, \bibinfo{author}{Zou, C.-L.} \&
  \bibinfo{author}{Guo, G.-C.}
\newblock \bibinfo{title}{Mechanical bound state in the continuum for
  optomechanical microresonators}.
\newblock \emph{\bibinfo{journal}{New Journal of Physics}}
  \textbf{\bibinfo{volume}{18}}, \bibinfo{pages}{063031}
  (\bibinfo{year}{2016}).

\bibitem{yu2021observation}
\bibinfo{author}{Yu, Y.}, \bibinfo{author}{Xi, X.} \& \bibinfo{author}{Sun, X.}
\newblock \bibinfo{title}{Observation of bound states in the continuum in a
  micromechanical resonator}.
\newblock \emph{\bibinfo{journal}{arXiv preprint arXiv:2109.09498}}
  (\bibinfo{year}{2021}).

\bibitem{zhao2019mechanical}
\bibinfo{author}{Zhao, M.} \& \bibinfo{author}{Fang, K.}
\newblock \bibinfo{title}{Mechanical bound states in the continuum for
  macroscopic optomechanics}.
\newblock \emph{\bibinfo{journal}{Optics Express}}
  \textbf{\bibinfo{volume}{27}}, \bibinfo{pages}{10138--10151}
  (\bibinfo{year}{2019}).

\bibitem{chan2012optimized}
\bibinfo{author}{Chan, J.}, \bibinfo{author}{Safavi-Naeini, A.~H.},
  \bibinfo{author}{Hill, J.~T.}, \bibinfo{author}{Meenehan, S.} \&
  \bibinfo{author}{Painter, O.}
\newblock \bibinfo{title}{Optimized optomechanical crystal cavity with acoustic
  radiation shield}.
\newblock \emph{\bibinfo{journal}{Applied Physics Letters}}
  \textbf{\bibinfo{volume}{101}}, \bibinfo{pages}{081115}
  (\bibinfo{year}{2012}).

\bibitem{fang2019anomalous}
\bibinfo{author}{Fang, K.} \& \bibinfo{author}{Wang, Y.}
\newblock \bibinfo{title}{Anomalous quantum hall effect of light in bloch-wave
  modulated photonic crystals}.
\newblock \emph{\bibinfo{journal}{Physical review letters}}
  \textbf{\bibinfo{volume}{122}}, \bibinfo{pages}{233904}
  (\bibinfo{year}{2019}).

\bibitem{snyder2013packaging}
\bibinfo{author}{Snyder, B.} \& \bibinfo{author}{O'Brien, P.}
\newblock \bibinfo{title}{Packaging process for grating-coupled silicon
  photonic waveguides using angle-polished fibers}.
\newblock \emph{\bibinfo{journal}{IEEE Transactions on Components, Packaging
  and Manufacturing Technology}} \textbf{\bibinfo{volume}{3}},
  \bibinfo{pages}{954--959} (\bibinfo{year}{2013}).

\bibitem{li2014silicon}
\bibinfo{author}{Li, C.}, \bibinfo{author}{Chee, K.~S.}, \bibinfo{author}{Tao,
  J.}, \bibinfo{author}{Zhang, H.}, \bibinfo{author}{Yu, M.} \&
  \bibinfo{author}{Lo, G.}
\newblock \bibinfo{title}{Silicon photonics packaging with lateral fiber
  coupling to apodized grating coupler embedded circuit}.
\newblock \emph{\bibinfo{journal}{Optics Express}}
  \textbf{\bibinfo{volume}{22}}, \bibinfo{pages}{24235--24240}
  (\bibinfo{year}{2014}).

\bibitem{hood2016atom}
\bibinfo{author}{Hood, J.~D.}, \bibinfo{author}{Goban, A.},
  \bibinfo{author}{Asenjo-Garcia, A.}, \bibinfo{author}{Lu, M.},
  \bibinfo{author}{Yu, S.-P.}, \bibinfo{author}{Chang, D.~E.} \&
  \bibinfo{author}{Kimble, H.}
\newblock \bibinfo{title}{Atom--atom interactions around the band edge of a
  photonic crystal waveguide}.
\newblock \emph{\bibinfo{journal}{Proceedings of the National Academy of
  Sciences}} \textbf{\bibinfo{volume}{113}}, \bibinfo{pages}{10507--10512}
  (\bibinfo{year}{2016}).

\bibitem{jin2019topologically}
\bibinfo{author}{Jin, J.}, \bibinfo{author}{Yin, X.}, \bibinfo{author}{Ni, L.},
  \bibinfo{author}{Solja{\v{c}}i{\'c}, M.}, \bibinfo{author}{Zhen, B.} \&
  \bibinfo{author}{Peng, C.}
\newblock \bibinfo{title}{Topologically enabled ultrahigh-{Q} guided resonances
  robust to out-of-plane scattering}.
\newblock \emph{\bibinfo{journal}{Nature}} \textbf{\bibinfo{volume}{574}},
  \bibinfo{pages}{501--504} (\bibinfo{year}{2019}).

\bibitem{chua2014larger}
\bibinfo{author}{Chua, S.-L.}, \bibinfo{author}{Lu, L.},
  \bibinfo{author}{Bravo-Abad, J.}, \bibinfo{author}{Joannopoulos, J.~D.} \&
  \bibinfo{author}{Solja{\v{c}}i{\'c}, M.}
\newblock \bibinfo{title}{Larger-area single-mode photonic crystal
  surface-emitting lasers enabled by an accidental dirac point}.
\newblock \emph{\bibinfo{journal}{Optics Letters}}
  \textbf{\bibinfo{volume}{39}}, \bibinfo{pages}{2072--2075}
  (\bibinfo{year}{2014}).

\bibitem{chen2022analytical}
\bibinfo{author}{Chen, Z.}, \bibinfo{author}{Yin, X.}, \bibinfo{author}{Li,
  P.}, \bibinfo{author}{Zheng, Z.}, \bibinfo{author}{Zhang, Z.},
  \bibinfo{author}{Wang, F.} \& \bibinfo{author}{Peng, C.}
\newblock \bibinfo{title}{Analytical theory of finite-size photonic crystal
  slabs near the band edge}.
\newblock \emph{\bibinfo{journal}{Optics Express}}
  \textbf{\bibinfo{volume}{30}}, \bibinfo{pages}{14033--14047}
  (\bibinfo{year}{2022}).

\bibitem{xu2005confined}
\bibinfo{author}{Xu, T.}, \bibinfo{author}{Yang, S.}, \bibinfo{author}{Nair,
  S.~V.} \& \bibinfo{author}{Ruda, H.}
\newblock \bibinfo{title}{Confined modes in finite-size photonic crystals}.
\newblock \emph{\bibinfo{journal}{Physical Review B}}
  \textbf{\bibinfo{volume}{72}}, \bibinfo{pages}{045126}
  (\bibinfo{year}{2005}).

\bibitem{regan2016direct}
\bibinfo{author}{Regan, E.~C.}, \bibinfo{author}{Igarashi, Y.},
  \bibinfo{author}{Zhen, B.}, \bibinfo{author}{Kaminer, I.},
  \bibinfo{author}{Hsu, C.~W.}, \bibinfo{author}{Shen, Y.},
  \bibinfo{author}{Joannopoulos, J.~D.} \& \bibinfo{author}{Solja{\v{c}}i{\'c},
  M.}
\newblock \bibinfo{title}{Direct imaging of isofrequency contours in photonic
  structures}.
\newblock \emph{\bibinfo{journal}{Science Advances}}
  \textbf{\bibinfo{volume}{2}}, \bibinfo{pages}{e1601591}
  (\bibinfo{year}{2016}).

\bibitem{ni2017analytical}
\bibinfo{author}{Ni, L.}, \bibinfo{author}{Jin, J.}, \bibinfo{author}{Peng, C.}
  \& \bibinfo{author}{Li, Z.}
\newblock \bibinfo{title}{Analytical and statistical investigation on
  structural fluctuations induced radiation in photonic crystal slabs}.
\newblock \emph{\bibinfo{journal}{Optics Express}}
  \textbf{\bibinfo{volume}{25}}, \bibinfo{pages}{5580--5593}
  (\bibinfo{year}{2017}).

\bibitem{nielsen2017multimode}
\bibinfo{author}{Nielsen, W. H.~P.}, \bibinfo{author}{Tsaturyan, Y.},
  \bibinfo{author}{M{\o}ller, C.~B.}, \bibinfo{author}{Polzik, E.~S.} \&
  \bibinfo{author}{Schliesser, A.}
\newblock \bibinfo{title}{Multimode optomechanical system in the quantum
  regime}.
\newblock \emph{\bibinfo{journal}{Proceedings of the National Academy of
  Sciences}} \textbf{\bibinfo{volume}{114}}, \bibinfo{pages}{62--66}
  (\bibinfo{year}{2017}).

\bibitem{renninger2018bulk}
\bibinfo{author}{Renninger, W.}, \bibinfo{author}{Kharel, P.},
  \bibinfo{author}{Behunin, R.} \& \bibinfo{author}{Rakich, P.}
\newblock \bibinfo{title}{Bulk crystalline optomechanics}.
\newblock \emph{\bibinfo{journal}{Nature Physics}}
  \textbf{\bibinfo{volume}{14}}, \bibinfo{pages}{601--607}
  (\bibinfo{year}{2018}).

\bibitem{zeller1971thermal}
\bibinfo{author}{Zeller, R.} \& \bibinfo{author}{Pohl, R.}
\newblock \bibinfo{title}{Thermal conductivity and specific heat of
  noncrystalline solids}.
\newblock \emph{\bibinfo{journal}{Physical Review B}}
  \textbf{\bibinfo{volume}{4}}, \bibinfo{pages}{2029} (\bibinfo{year}{1971}).

\bibitem{thompson1961thermal}
\bibinfo{author}{Thompson, J.} \& \bibinfo{author}{Younglove, B.}
\newblock \bibinfo{title}{Thermal conductivity of silicon at low temperatures}.
\newblock \emph{\bibinfo{journal}{Journal of Physics and Chemistry of Solids}}
  \textbf{\bibinfo{volume}{20}}, \bibinfo{pages}{146--149}
  (\bibinfo{year}{1961}).

\bibitem{chu2017quantum}
\bibinfo{author}{Chu, Y.}, \bibinfo{author}{Kharel, P.},
  \bibinfo{author}{Renninger, W.~H.}, \bibinfo{author}{Burkhart, L.~D.},
  \bibinfo{author}{Frunzio, L.}, \bibinfo{author}{Rakich, P.~T.} \&
  \bibinfo{author}{Schoelkopf, R.~J.}
\newblock \bibinfo{title}{Quantum acoustics with superconducting qubits}.
\newblock \emph{\bibinfo{journal}{Science}} \textbf{\bibinfo{volume}{358}},
  \bibinfo{pages}{199--202} (\bibinfo{year}{2017}).

\bibitem{kotler2021direct}
\bibinfo{author}{Kotler, S.}, \bibinfo{author}{Peterson, G.~A.},
  \bibinfo{author}{Shojaee, E.}, \bibinfo{author}{Lecocq, F.},
  \bibinfo{author}{Cicak, K.}, \bibinfo{author}{Kwiatkowski, A.},
  \bibinfo{author}{Geller, S.}, \bibinfo{author}{Glancy, S.},
  \bibinfo{author}{Knill, E.}, \bibinfo{author}{Simmonds, R.~W.} \emph{et~al.}
\newblock \bibinfo{title}{Direct observation of deterministic macroscopic
  entanglement}.
\newblock \emph{\bibinfo{journal}{Science}} \textbf{\bibinfo{volume}{372}},
  \bibinfo{pages}{622--625} (\bibinfo{year}{2021}).

\bibitem{safavi2013laser}
\bibinfo{author}{Safavi-Naeini, A.~H.}, \bibinfo{author}{Chan, J.},
  \bibinfo{author}{Hill, J.~T.}, \bibinfo{author}{Gr{\"o}blacher, S.},
  \bibinfo{author}{Miao, H.}, \bibinfo{author}{Chen, Y.},
  \bibinfo{author}{Aspelmeyer, M.} \& \bibinfo{author}{Painter, O.}
\newblock \bibinfo{title}{Laser noise in cavity-optomechanical cooling and
  thermometry}.
\newblock \emph{\bibinfo{journal}{New Journal of Physics}}
  \textbf{\bibinfo{volume}{15}}, \bibinfo{pages}{035007}
  (\bibinfo{year}{2013}).
  
\bibitem{meenehan2014silicon}
\bibinfo{author}{Meenehan, S.~M.}, \bibinfo{author}{Cohen, J.~D.},
  \bibinfo{author}{Gr{\"o}blacher, S.}, \bibinfo{author}{Hill, J.~T.},
  \bibinfo{author}{Safavi-Naeini, A.~H.}, \bibinfo{author}{Aspelmeyer, M.} \&
  \bibinfo{author}{Painter, O.}
\newblock \bibinfo{title}{Silicon optomechanical crystal resonator at
  millikelvin temperatures}.
\newblock \emph{\bibinfo{journal}{Physical Review A}}
  \textbf{\bibinfo{volume}{90}}, \bibinfo{pages}{011803}
  (\bibinfo{year}{2014}).

\bibitem{safavi2013laser}
\bibinfo{author}{Safavi-Naeini, A.~H.}, \bibinfo{author}{Chan, J.},
  \bibinfo{author}{Hill, J.~T.}, \bibinfo{author}{Gr{\"o}blacher, S.},
  \bibinfo{author}{Miao, H.}, \bibinfo{author}{Chen, Y.},
  \bibinfo{author}{Aspelmeyer, M.} \& \bibinfo{author}{Painter, O.}
\newblock \bibinfo{title}{Laser noise in cavity-optomechanical cooling and
  thermometry}.
\newblock \emph{\bibinfo{journal}{New Journal of Physics}}
  \textbf{\bibinfo{volume}{15}}, \bibinfo{pages}{035007}
  (\bibinfo{year}{2013}).

\bibitem{jin2019topologically}
\bibinfo{author}{Jin, J.}, \bibinfo{author}{Yin, X.}, \bibinfo{author}{Ni, L.},
  \bibinfo{author}{Solja{\v{c}}i{\'c}, M.}, \bibinfo{author}{Zhen, B.} \&
  \bibinfo{author}{Peng, C.}
\newblock \bibinfo{title}{Topologically enabled ultrahigh-q guided resonances
  robust to out-of-plane scattering}.
\newblock \emph{\bibinfo{journal}{Nature}} \textbf{\bibinfo{volume}{574}},
  \bibinfo{pages}{501--504} (\bibinfo{year}{2019}).

\bibitem{zhao2020design}
\bibinfo{author}{Zhao, Z.} \& \bibinfo{author}{Fan, S.}
\newblock \bibinfo{title}{Design principles of apodized grating couplers}.
\newblock \emph{\bibinfo{journal}{Journal of Lightwave Technology}}
  \textbf{\bibinfo{volume}{38}}, \bibinfo{pages}{4435--4446}
  (\bibinfo{year}{2020}).

\bibitem{molesky2018inverse}
\bibinfo{author}{Molesky, S.}, \bibinfo{author}{Lin, Z.},
  \bibinfo{author}{Piggott, A.~Y.}, \bibinfo{author}{Jin, W.},
  \bibinfo{author}{Vuckovi{\'c}, J.} \& \bibinfo{author}{Rodriguez, A.~W.}
\newblock \bibinfo{title}{Inverse design in nanophotonics}.
\newblock \emph{\bibinfo{journal}{Nature Photonics}}
  \textbf{\bibinfo{volume}{12}}, \bibinfo{pages}{659--670}
  (\bibinfo{year}{2018}).

\bibitem{akahane2005fine}
\bibinfo{author}{Akahane, Y.}, \bibinfo{author}{Asano, T.},
  \bibinfo{author}{Song, B.-S.} \& \bibinfo{author}{Noda, S.}
\newblock \bibinfo{title}{Fine-tuned high-{Q} photonic-crystal nanocavity}.
\newblock \emph{\bibinfo{journal}{Optics Express}}
  \textbf{\bibinfo{volume}{13}}, \bibinfo{pages}{1202--1214}
  (\bibinfo{year}{2005}).  
  

\end{thebibliography}
\end{document}